
\magnification\magstep1
\baselineskip14pt
\vsize23.5truecm 
\hsize16.5truecm




\def\hatt{\widehat}
\def\dell{\partial}
\def\tilda{\widetilde}
\def\eps{\varepsilon}

\def\half{\hbox{$1\over2$}}

\def\quart{\hbox{$1\over4$}}

\def\arr{\rightarrow}
\def\normal{{\cal N}}
\def\RR{\mathord{I\kern-.3em R}}
\def\PP{\mathord{I\kern-.3em P}}
\def\NN{\mathord{I\kern-.3em N}}
\def\ZZ{\mathord{I\kern-.3em Z}} 
\def\Var{{\rm Var}}
\def\E{{\rm E}}
\def\d{{\rm d}}

\def\midd{{\,|\,}}
\def\subsection{\medskip}

\font\bigbf=cmbx12

\font\csc=cmcsc10

 at 10truept 
\font\smallrm=cmr8

\def\today{\number\day \space \ifcase\month\or
January\or February\or March\or April\or May\or June\or 
July\or August\or September\or October\or November\or December\fi  
\space \number\year}


   
\def\ref#1{{\noindent\hangafter=1\hangindent=20pt
  #1\smallskip}}   	  

\def\quotationone{\smallrm Where there is a Will}
\def\quotationtwo{\smallrm There is a Won't}
\def\hskipdistanceleft{\hskip-3.8pt}
\def\hskipdistanceright{\hskip-2.0pt}

\footline={{
\ifodd\count0
	{\hskipdistanceleft\quotationone\phantom{\today\hskip-30pt}
		\hfil{\rm\the\pageno}\hfil
	 \phantom{\quotationone\hskip-30pt}{\smallrm\today}\hskipdistanceright}
	\else 
	{\hskipdistanceleft\quotationtwo\phantom{\today\hskip-30pt}
		\hfil{\rm\the\pageno}\hfil
	 \phantom{\quotationtwo\hskip-30pt}{\smallrm\today}\hskipdistanceright}
	\fi}}


\def\cstok#1{\leavevmode\thinspace\hbox{\vrule\vtop{\vbox{\hrule\kern1pt
        \hbox{\vphantom{\tt/}\thinspace{\tt#1}\thinspace}}
        \kern1pt\hrule}\vrule}\thinspace} 
\def\square{\cstok{\phantom{$\cdot$}}} 


\def\fermat#1{\setbox0=\vtop{\hsize5.00pc
	\smallrm\raggedright\noindent\baselineskip9pt
	\rightskip=0.5pc plus 1.5pc #1}\leavevmode
	\vadjust{\dimen0=\dp0
	\kern-\ht0\hbox{\kern-5.00pc\box0}\kern-\dimen0}}


\def\sumin{\sum_{i=1}^n}
\def\de{\delta}
\def\ga{\gamma}
\def\twentyfourth{\hbox{$1\over24$}}
\def\rootpi{\sqrt{\pi}}

\def\today{{January 1994}}

\def\quotationone{\smallrm{Hjort Glad }}
\def\quotationtwo{\smallrm Semiparametric density estimation} 

\centerline{\bigbf Nonparametric density estimation with a parametric start}

\medskip
\centerline{\bf Nils Lid Hjort$^1$ and Ingrid K.~Glad$^2$}

\smallskip
\centerline{\bf $^1$University of Oslo and}
\centerline{\bf $^2$The Norwegian Institute of Technology, Trondheim}

\smallskip
{{\smallskip\narrower\noindent\baselineskip12pt
{\csc Abstract.} 
The traditional kernel density estimator of an unknown density
is by construction completely nonparametric, in the sense that
it has no preferences and will work reasonably well for all shapes. 
The present paper develops a class of semiparametric methods that 
are designed to work better than the kernel estimator 
in a broad nonparametric neighbourhood 
of a given parametric class of densities, for example the normal, 
while not losing much in precision when the true density is far from the
parametric class. 
The idea is to multiply an initial parametric density estimate 
with a kernel type estimate of the necessary correction factor. 
This works well in cases where the correction factor 
function is less rough than the original density itself. 
Extensive comparisons with the kernel estimator are carried out,
including exact analysis for the class of all normal mixtures.
The new method, with a normal start, wins quite often,
even in many cases where the true density is far from normal.   
Procedures for choosing the smoothing parameter of the estimator
are also discussed.  
The new estimator should be particularly useful in higher dimensions,  
where the usual nonparametric methods have problems. 
The idea is also spelled out for nonparametric regression. 

\smallskip\noindent
{\csc Key words:} \sl
bandwidth selection, 
correction factor,  
kernel methods, 
lowering the bias, 
semiparametric density estimation,
test cases 
\smallskip}} 

\bigskip
{\bf 1. Introduction and summary.}
Let $X_1,\ldots,X_n$ be independent observations from an unknown density $f$ 
on the real line. The traditional nonparametric density estimator is 
$$\tilda f(x)={1\over n}\sumin h^{-1}K(h^{-1}(X_i-x))
	={1\over n}\sumin K_h(X_i-x), \eqno(1.1)$$
where $K_h(z)=h^{-1}K(h^{-1}z)$ and $K(z)$ is a kernel function,
which is taken here to be a symmetric probability density with finite
values of $\sigma_K^2=\int z^2K(z)\,\d z$ and $R(K)=\int K(z)^2\,\d z$. 
The basic statistical properties are that
$$\E\tilda f(x)\doteq f(x)+\half\sigma_K^2h^2f''(x)
\quad {\rm and} \quad
\Var\,\tilda f(x)\doteq R(K)(nh)^{-1}\,f(x)-f(x)^2/n. \eqno(1.2)$$
The integrated mean squared error is of order $n^{-4/5}$ when 
$h$ is proportional to $n^{-1/5}$, which is the optimal size.
See Scott (1992, Chapter 6) and Wand \& Jones (1994, Chapter 2) 
for recent accounts of the theory. 

Method (1.1) is totally nonparametric and admirably impartial to 
special types of shapes of the underlying density. 
The intention of the present paper is to construct competitors 
to (1.1) with properties that are generally similar 
but indeed better in the broad vicinity of given parametric families. 
The basic idea is to start out with a parametric density estimate
$f(x,\hatt\theta)$, say the normal, and then multiply with
a nonparametric kernel type estimate of the correction function 
$r(x)=f(x)/f(x,\hatt\theta)$. Our proposal is 
$\hatt r(x)=n^{-1}\sumin K_h(X_i-x)/f(X_i,\hatt\theta)$, producing
$$\hatt f(x)=f(x,\hatt\theta)\hatt r(x)
	={1\over n}\sumin K_h(X_i-x)
	{f(x,\hatt\theta)\over f(X_i,\hatt\theta)}. \eqno(1.3)$$
We emphasise that the initial parametric estimate is not 
(necessarily) intended to provide a serious approximation 
to the true density; our method will often work well even if the
parametric description is quite crude. The case of a constant
start value for $f(x,\theta)$, corresponding to choosing a 
uniform distribution as the initial description, gives back the
classic kernel estimator (1.1). 

The basic bias and variance properties of the new estimator (1.3) are 
investigated in Section 2, treating the simplest case of
a non-random start function $f_0(x)$, and in Section 3, 
covering a broad class of parametric start estimators. 
It turns out that the variance of the (1.3) estimator 
is simply the same as the variance of the traditional (1.1) estimator,
to the order of approximation used, while the bias is 
quite similar in structure to (1.2), and often smaller. 
Comparisons with the traditional estimator (1.1) are made 
in Sections 4 and 5. 
It is seen that the new method generally is the better one 
in cases where the correction function 
is less `rough' than the original density,
in a sense made precise in Section 4, and illustrated there 
in the realm of Hermite expansions around the normal.  

Further analysis is provided in Section 5, 
for the version of (1.3) that starts with the normal,
comparing behaviour with the kernel method 
when the true density belongs to the large class of all normal mixtures. 
There and in the paper's appendix comparative formulae are developed 
for exact analysis of asymptotic mean squared error 
as well as for exact finite-sample mean squared error.  
The results are illuminated by working through 
a list of 15 `test densities' proposed by Marron \& Wand (1992), 
chosen to exhibit a broad range of distributional shapes. 
The new `nonparametrically corrected normal estimate' 
outperforms the usual kernel method in 12 of these 15 test cases,
and in all the `not drastically unreasonable' cases,
in terms of approximate mean integrated squared error.  
The same pattern is observed for finite sample sizes. 

The bottom line is that (1.3) will be more precise than (1.1) 
in a broad nonparametric neighbourhood around the parametric family,
while at the same time losing surprisingly little, or not at all, 
when the true density is far from the parametric family. 
One explanation is that the uniform prior description, 
which in the light of (1.3) is the implicit start estimator 
for the kernel estimator (1.1), 
is overly conservative and less advantageous than 
say the normal, even in quite non-normal cases. 

The problem of selecting a good smoothing parameter 
is discussed in Section 6, and some solutions are outlined,
including versions of plug-in and cross validation.  
Our method also works well in the multi-dimensional case, 
starting out for example with a multi-normal start estimate, 
as demonstrated in Section 7. The method should be particularly 
useful in the higher-dimensional case since the ordinary 
nonparametric methods, including the kernel method, are 
quite imprecise then. 
Our paper ends with some supplementary comments in Section 8.
In particular Remark 8E spells out the corresponding 
estimation idea for nonparametric regression, 
giving a generalised Nadaraya--Watson estimator. 

Our estimators can be viewed as semiparametric in that 
they combine parametric and nonparametric methods. 
They are as such in the same realm as 
recent methods of Hjort (1993) and Hjort \& Jones (1993).
These latter methods are quite different but 
also have the property that the variance is approximately the same 
as in (1.2) while the bias is similar but sometimes smaller. 
The (1.3) method is also similar in spirit to the projection pursuit
density estimation methods, see for example 
Friedman, Stuetzle \& Schroeder (1984), 
and also to the normal times Hermite expansion method, 
see for example Hjort (1986), Buckland (1992),
and Hjort \& Fenstad (1994).  
A somewhat less attractive semiparametric method is
that of Schuster \& Yakowitz (1985) and Olkin \& Spiegelman (1987), 
see the discussion in Jones (1993). 
Various semiparametric Bayesian density estimators are proposed 
in Hjort (1994). 

Another semiparametric technique, perhaps mildly related to our new method, 
is the transformation idea of Wand, Marron \& Ruppert (1991), 
where data are semiparametrically transformed so as to work well with 
a non-adaptive constant smoothing parameter, 
and then ending in a back-transformed density estimator. 
This is a promising way of using an adaptive smoothing parameter, 
and our estimator can be seen as as having similar intentions. 
In other words, (1.3) can be seen as being similar in spirit 
to a suitable semiparametrically adaptive 
$n^{-1}\sumin K_{h(x,\hatt\theta)}(X_i-x)$. 
Finally we mention a recent bias reduction method due to 
Jones, Linton \& Nielsen (1993). Our (1.3) idea is to 
start with any parametric estimator and then multiply 
with a nonparametric correction function, and in essence 
this does not affect the variance but changes the bias. 
Serendipitously and independently of the present authors 
Jones, Linton \& Nielsen (1993) use essentially the 
same idea but in a totally nonparametric mode, 
correcting the initial kernel estimator with a nonparametric 
correction factor in the (3.1) manner. 
This typically gives a smaller bias but a somewhat larger variance. 

\bigskip
{\bf 2. Nonparametric correction on a fixed start.}
Suppose $f_0$ is a fixed density, perhaps a crude guess of $f$. 
Write $f=f_0r$. The idea is to estimate the nonparametric correction 
factor $r$ via kernel smoothing. One version of this is 
$\hatt r(x)=n^{-1}\sumin K_h(X_i-x)/f_0(X_i)$, with ensuing estimator
$$\hatt f(x)=f_0(x)\hatt r(x)
	={1\over n}\sumin K_h(X_i-x){f_0(x)\over f_0(X_i)}. \eqno(2.1)$$
Note that a constant $f_0(x)$ gives back the ordinary kernel estimator (1.1).  
We have
$$\eqalign{
\E\hatt r(x)&=\int K_h(y-x)f_0(y)^{-1}f(y)\,\d y \cr
	&=\int K(z)r(x+hz)\,\d z
	 =r(x)+\half\sigma_K^2h^2r''(x)+O(h^4), \cr}$$
and 
$$\eqalign{\Var\,\hatt r(x)
&={1\over n}\Bigl[\int {K_h(y-x)^2\over f_0(y)^2}f(y)\,\d y	
	-\{\E\hatt r(x)\}^2\Bigr] \cr
&={R(K)\over nh}{f(x)\over f_0(x)^2}-{r(x)^2\over n}+O(h/n), \cr}$$
by a variation of the arguments traditionally used to establish (1.2).
This shows that the (2.1) estimator has
$${\rm bias}\doteq \half\sigma_K^2h^2f_0(x)r''(x)
\quad{\rm and} \quad
{\rm variance}\doteq R(K)(nh)^{-1}\,f(x)-f(x)^2/n. \eqno(2.2)$$
In other words, the variance is of the very same size as that of 
the traditional estimator, to the order of approximation used, 
and the bias is of the same order $h^2$, 
but proportional to $f_0r''$ rather than to $f''$.
The new estimator is better than the traditional one in 
all cases where $f_0r''$ is smaller in size than 
$f''=f_0''r+2f_0'r'+f_0r''$. In cases where $f_0$ is already 
a good guess one expects $r$ near constant and $r''$ small,
so this describes a certain neighbourhood of densities 
around $f_0$ where the new method is better than the traditional one.
This is further discussed and exemplified in Section 4.

\bigskip
{\bf 3. Nonparametric correction on a parametric start.}
Let $f(x,\theta)$ be a given parametric family of densities,
where the possibly multi-dimensional parameter 
$\theta=(\theta_1,\allowbreak\ldots,\theta_p)'$ belongs to some open and
connected region in $p$-space. 
The parametric start estimate is $f(x,\hatt\theta)$, 
where we for concreteness let $\hatt\theta$ be the maximum likelihood
estimator (quite general estimators for $\theta$ are allowed later). 
Thus $f(x,\hatt\theta)$ could be the 
estimated normal density, for example, or an estimated mixture
of two normals. This initial data summary is 
not necessarily meant to be a serious description of the true density; 
the method we will develop is intended to work well
even if $f$ cannot be well approximated by any $f(.,\theta)$. 

The task is to estimate the necessary correction function 
$f(x)/f(x,\hatt\theta)$ by kernel smoothing means. 
In view of Section 2 
$\hatt r(x)=n^{-1}\sumin K_h(X_i-x)/f(X_i,\hatt\theta)$
is a natural choice. 
In other words,
$$\hatt f(x)=f(x,\hatt\theta){1\over n}\sumin 
	K_h(X_i-x)/f(X_i,\hatt\theta). \eqno(3.1)$$

In order to understand to what extent the parametric estimation
makes this estimator quantitatively different from the cleaner 
version (2.1), we bring in facts about the behaviour of the 
maximum likelihood estimator outside model conditions. 
It aims at a certain $\theta_0$,
the least false value according to the Kullback--Leibler distance
measure $\int f(x)\log\{f(x)/f(x,\theta)\}\,\d x$ 
from true $f$ to approximant $f(.,\theta)$. Write 
$f_0(x)=f(x,\theta_0)$ for this best parametric approximant,
and let $u_0(x)=\dell\log f(x,\theta_0)/\dell\theta$ be the score
function evaluated at this parameter value. A Taylor expansion gives
$$\eqalign{{f(x,\hatt\theta)\over f(X_i,\hatt\theta)}
&=\exp\{\log f(x,\hatt\theta)-\log f(X_i,\hatt\theta)\} \cr
&\doteq{f_0(x)\over f_0(X_i)}+{f_0(x)\over f_0(X_i)}\{u_0(x)-u_0(X_i)\}'
	(\hatt\theta-\theta_0), \cr} \eqno(3.2)$$
leading to 
$$\eqalign{\hatt f(x)
&\doteq {1\over n}\sumin K_h(X_i-x){f_0(x)\over f_0(X_i)}
	\bigl[1-\{u_0(X_i)-u_0(x)\}'(\hatt\theta-\theta_0)\bigr] \cr
&=f^*(x)+V_n(x), \cr}\eqno(3.3)$$
say. Here $f^*$ is as in (2.1), except for the fact that the $f_0$ 
function appearing here is not directly visible, and the $V_n(x)$ term
stems from the parametric estimation variability. 

Representation (3.3), in concert with expressing $\hatt\theta-\theta_0$
as an average of i.i.d.~zero mean variables plus remainder term,
can now be used to establish approximate bias and variance results for
$\hatt f(x)$. We shall be somewhat more general and allow arbitrary
regular estimators having an influence with finite covariance matrix. 
To define this properly, let $F$ be the true distribution, the 
cumulative of $f$, and let $F_n$ be the empirical distribution function.
We consider functional estimators of $\theta$ of the form  
$\hatt\theta=T(F_n)$ with influence function 
$I(x)=\lim_{\eps\arr0}\{T((1-\eps)F+\eps\delta_x)-T(F)\}/\eps$,
writing $\delta_x$ for unit point mass at $x$, 
and assume that $\Sigma=\E_fI(X_i)I(X_i)'$ is finite. 
The best approximant $f_0(x)=f(x,\theta_0)$ to $f(x)$ that 
$f(x,\hatt\theta)$ aims for is determined by $\theta_0=T(F)$.  
Under mild regularity conditions, 
see for example Huber (1981) or Shao (1991), one has  
$$\hatt\theta-\theta_0={1\over n}\sumin I(X_i)+{d\over n}+\eps_n, \eqno(3.4)$$
where $\eps_n=O_p(n^{-1})$ with mean $O(n^{-2})$, 
i.e.~$d/n$ is essentially the bias of $\hatt\theta$. 
It is generally possible to de-bias the estimator, 
for example by jackknifing or bootstrapping, making the $d/n$ term
disappear.  
The maximum likelihood case corresponds to $I(x)=J^{-1}u_0(x)$ where 
$J=-\E_f\dell^2\log f(X_i,\theta_0)/\dell\theta\dell\theta'$. 

{{\smallskip\sl
{\csc Proposition.}
Let $f_0(x)=f(x,\theta_0)$ with $\theta_0=T(F)$ 
be the best parametric approximant to $f$, and let $r=f/f_0$. 
The semiparametric estimator (3.1) has 
$$\eqalign{
\E\hatt f(x)&=f(x)+\half\sigma_K^2h^2f_0(x)r''(x)+O(h^2/n+h^4+n^{-2}) \cr
{\sl and}\quad\Var\,\hatt f(x)&=R(K)(nh)^{-1}\,f(x)
	-f(x)^2/n+O(h/n+n^{-2}). \cr}$$
\smallskip}}

{\csc Proof:} 
The detailed proof we present needs a second order Taylor 
approximation version of the simpler first order Taylor versions 
(3.2)--(3.3). This more complete approximation becomes 
$\hatt f(x)=f^*(x)+V_n(x)+\half W_n(x)$, where we write 
$f^*(x)=\bar A_n$,
$V_n(x)=\bar B_n'(\hatt\theta-\theta_0)$, and
$W_n(x)=(\hatt\theta-\theta_0)'\bar C_n(\hatt\theta-\theta_0)$.
The representations are in terms of averages of i.i.d.~variables 
$$\eqalign{
A_i&=K_h(X_i-x)f_0(x)/f_0(X_i), \cr
B_i&=-K_h(X_i-x)\{f_0(x)/f_0(X_i)\}\,\{u_0(X_i)-u_0(x)\}, \cr
C_i&=K_h(X_i-x)\{f_0(x)/f_0(X_i)\}\,w(x,X_i), \cr}$$
where in fact 
$w(x,X_i)=v_0(x)-v_0(X_i)+\{u_0(x)-u_0(X_i)\}\{u_0(x)-u_0(X_i)\}'$. 

Starting with the expected value, we already know that
$$ f^* \quad {\rm has\ mean}\quad f(x)+\half\sigma_K^2h^2f_0(x)r''(x). $$
Through (3.4) and the averages representations above one finds 
$\E V_n(x)=n^{-1}\E B_i'I_i+n^{-1}(\E B_i)'d+O(n^{-2})$ 
and $\E W_n(x)=n^{-1}{\rm Tr}(\E C_i\,\E I_iI_i')+O(n^{-2})$,   
using the fact that $I_i=I(X_i)$ has mean zero. 
But it is not difficult to see that each of 
$\E B_i$, $\E B_i'I_i$, $\E C_i$ is of size $O(h^2)$; for example, 
$$\eqalign{
\E B_i'I_i&=-\int K_h(y-x){f_0(x)\over f_0(y)}\{u_0(y)-u_0(x)\}'
	I(y)f(y)\,d y \cr
&=-\int K(z)f_0(x)\{u_0(x+hz)-u_0(x)\}'(Ir)(x+hz)\,\d z \cr
&=-h^2\sigma_K^2f_0(x)\{u_0'(x)'(Ir)'(x)+\half u_0''(x)'(Ir)(x)\}
	+O(h^4). \cr}$$
One can also see that the remainder of the second order Taylor
approximation used, involving $(\hatt\theta_i-\theta_{0,i})^3$
terms, is of size $O_p(n^{-2})$.
Thus the bias of $\hatt f(x)$ is 
$\half\sigma_K^2h^2f_0(x)r''(x)+(h^2/n)b(x)+O(h^4+n^{-2})$,
for a certain $b(x)$ function. 

Next turn to the variance. The variance of $f^*(x)$ is known from
Section 2. From (3.4) and the representation above one finds
$$\eqalign{
\Var\,V_n(x)
&=\Var(\bar B_n'\bar I_n)+O(n^{-2}) \cr
&=n^{-1}(\E B_i)'\Sigma(\E B_i)-\{O(h^2/n)\}^2+O(n^{-2})=O(h^4/n+n^{-2}), }$$
and similarly $W_n(x)$ can be seen to have 
uninfluential variance $O(h^4/n^2)$. 
Finally
$$ {\rm cov}\{f^*(x),V_n(x)\}=n^{-1}(\E B_i)'\E A_iI_i+O(n^{-2}) =O(h^2/n). $$
This combines to give the necessary variance expression. 
\square 

\smallskip
The result is remarkable in its simplicity; 
the sizes of bias and variance are only affected by 
parametric estimation noise to the quite small $O(h^2/n+n^{-2})$ order.  
The reason lies with (3.2); 
not only is $\hatt\theta$ close to $\theta_0$, 
but the $\hatt f(x)$ estimator uses only $X_i$s that are close to $x$,
making $u_0(X_i)$ close to $u_0(x)$. 
The story is somewhat different for the correction term 
$\hatt r(x)$ alone, see Remark 8B. 

Consistency of the density estimator requires both 
$h\arr0$ (forcing the bias towards zero) and
$nh\arr\infty$ (making the variance go to zero). 
The optimal size of $h$ will later be seen to be 
proportional to $n^{-1/5}$. These observations match 
the traditional facts for the classic (1.1) estimator.   
Note also that if the parametric model happens to be 
accurate, then the $r$ function is equal to 1, and 
the bias is only $O(h^4+h^2/n)$. 

\smallskip
{\csc Example 1: Normal start estimate.}
The normal start estimate is of the form 
$\hatt\sigma^{-1}\phi(\hatt\sigma^{-1}(x-\hatt\mu))$,
where one can use maximum likelihood estimates 
$\hatt\mu=n^{-1}\sumin X_i$ and 
$\hatt\sigma^2=n^{-1}\sumin(X_i-\hatt\mu)^2$ (or the de-biased 
version with denominator $n-1$).  
In view of the generality of the proposition above quite general
estimators are allowed, without changing the basic structure of
bias and variance of $\hatt f(x)$. One might for example wish 
to use robust estimates of mean and standard deviation.     
In any case the density estimator is 
$$\eqalign{\hatt f(x)
&={1\over \hatt\sigma}\phi\Bigl({x-\hatt\mu\over \hatt\sigma}\Bigr)
	{1\over n}\sumin K_h(X_i-x)\Big/
	{1\over \hatt\sigma}\phi\Bigl({X_i-\hatt\mu\over \hatt\sigma}\Bigr) \cr
&={1\over n}\sumin K_h(X_i-x){\exp\{-\half(x-\hatt\mu)^2/\hatt\sigma^2\}
	\over \exp\{-\half(X_i-\hatt\mu)^2/\hatt\sigma^2\}}. \cr}\eqno(3.5)$$
Note that its implementation is straightforward. 

\smallskip
{\csc Example 2: Log-normal start estimate.} 
One option for positive data is to start with a log-normal
approximation and then multiply with a correction factor. The result is
$$\hatt f(x)={1\over n}\sumin K_h(X_i-x)
	{\exp\{-\half(\log x-\hatt\mu)^2/\hatt\sigma^2\}
	\over \exp\{-\half(\log X_i-\hatt\mu)^2/\hatt\sigma^2\}}
	{X_i\over x}.$$

\smallskip
{\csc Example 3: Gamma start estimate.}
A version of the general method which should work well for
positive data from perhaps unimodal and right-skewed distributions 
is to start with a gamma distribution approximation.  
The final estimator is then of the form 
$$\hatt f(x)={1\over n}\sumin K_h(X_i-x)
	(x/X_i)^{\hatt\alpha-1}\exp\{-\hatt\beta(x-X_i)\},$$
for example with moment estimates for the gamma parameters. 

\smallskip
{\csc Example 4: Normal mixture start estimate.}
We believe proper use of the three special cases mentioned 
now would work satisfactorily in many applications. 
Most unimodal densities would be approximable with either
a normal, a log-normal or a gamma, perhaps after a transformation.   
Cases where still other tactics might prove superior include 
densities exhibiting two or more bumps. One method in such
cases would be to fit a normal mixture first and use that as 
the $f(x,\hatt\theta)$, correcting afterwards with a $\hatt r(x)$. 

\smallskip
{\csc Remark 1.} 
The correction factor $f(x,\hatt\theta)/f(X_i,\hatt\theta)$ 
can occasionally be too influential, in cases 
where the denominator is too small. This is not a problem for 
small $h$ since then only $X_i$s quite close to $x$ matter,
but it can happen for moderate $h$ and for lonely data points. 
An effective safety procedure is to replace 
the original $f(x,\hatt\theta)$ function with a somewhat 
adjusted $\bar f(x,\hatt\theta)$, bounding it suitably away from zero.
In the normal case we advocate putting $\bar f(x,\theta)$ equal to 
$f(\hatt\mu\pm2.5\,\hatt\sigma,\hatt\mu,\hatt\sigma)
=(2\pi)^{-1/2}\hatt\sigma^{-1}\exp(-2.5^2/2)$ 
for $|x-\hatt\mu|\ge2.5\,\hatt\sigma$. 

\smallskip
{\csc Remark 2.} 
We have developed a method that can be used for any given parametric
model. It is intuitively clear that the method works best in cases
where the model employed is not too far from covering the truth
(and this is borne out by precise analysis in the following sections). 
One could think of ways of automatising 
the choice of the parametric vehicle model, 
through suitable goodness of fit measures,
thereby obtaining an overall adaptive density estimator,
but this is not pursued here.

\bigskip
{\bf 4. Comparison with the traditional kernel density estimator.}
In this and the following section the performance of 
the new estimator is compared to that of the usual (1.1) estimator. 
We look into a couple of `test areas', that is, classes of densities 
for which comparison of behaviour can be carried out. 
In 4B and 4C below we study two versions of Hermite expansions 
around the normal density. 
The calculations we give for these
turn out to be useful also in connection with the problem
of choosing the bandwidth parameter $h$, see Section 6.  
The second test area is that of finite normal mixtures, 
studied in Section 5 and in the Appendix, 
with attention given to the list of 15 test densities 
chosen by Marron \& Wand (1992). 

\subsection
{\csc 4A. General mse and mise comparison.}
Expressions can be found for the leading terms of the
integrated mean squared errors of the usual kernel estimator (1.1)
and the new estimator (3.1), using respectively (1.2) and 
the proposition of Section 3. We find 
$$\eqalign{
{\rm amise\ for\ }\tilda f&=\quart\sigma_K^4h^4R_{\rm trad}(f)
	+R(K)(nh)^{-1}, \cr
{\rm amise\ for\ }\hatt f&=\quart\sigma_K^4h^4R_{\rm new}(f)
	+R(K)(nh)^{-1}, \cr} \eqno(4.1)$$
featuring `roughness' functionals 
$$R_{\rm trad}(f)=\int \{f''(x)\}^2\,\d x
\quad{\rm and} \quad
R_{\rm new}(f)=\int \{f_0(x)r''(x)\}^2\,\d x. \eqno(4.2)$$
The new estimator is better, in the sense of approximate (leading terms)
integrated mean squared error, whenever 
$R_{\rm new}(f)$ is smaller than $R_{\rm trad}(f)$. 
This defines a nonparametric neighbourhood of 
densities around the parametric class. 	
When $f$ belongs to this neighbourhood, 
$\hatt f$ is better than $\tilda f$ when the same $K$ and the same $h$
are used in the two estimators. In such a case the new estimator
can be made even better by choosing an appropriate $h$, 
see Section 6.

It is also of interest to see in which $x$-regions the new estimator
is better than the traditional one. 
Write $f=\exp(g)$ and $f_0=\exp(g_0)$. Then 
$$f''=f\{g''+(g')^2\} \quad {\rm while} \quad 
  f_0r''=f\{g''-g_0''+(g'-g_0')^2\}. \eqno(4.3) $$ 
This is useful for actual inspection of the bias terms for different $f$s,
and is attractive in that it clearly exhibits the roles of 
the first and second log-derivatives. Note in particular that 
if the parametric model used is good enough to secure 
$|g'-g_0'|\le|g'|$ and $|g''-g_0''|\le|g''|$, for a region of 
relevant $x$s, then that clearly suffices for the new method
to be better than the traditional one. 
These requirements can also be written 
$0\le g_0'/g'\le2$ and $0\le g_0''/g''\le2$. 

\subsection
{\csc 4B. First test-bed: Hermite expansions.}
A test area where these matters can be explored 
is in the context of the Hermite expansions 
considered (for other purposes) in Hjort \& Jones (1994) 
and in Hjort \& Fenstad (1994).
Let $H_j(x)$ be the $j$th Hermite polynomial, given by
$\phi^{(j)}(x)=(-1)^j\phi(x)\allowbreak H_j(x)$. 
Consider the Hermite expansion representation 
$$f(x)=\phi\Bigl({x-\mu\over \sigma}\Bigr){1\over \sigma}
	\Bigl\{1+\sum_{j=3}^m{\ga_j\over j!}
	H_j\Bigl({x-\mu\over \sigma}\Bigr)\Bigr\}
	=g\Bigl({x-\mu\over \sigma}\Bigr){1\over \sigma}, \eqno(4.4)$$
writing $g(y)=\phi(y)\{1+\sum_{j=3}^m(\ga_j/j!)H_j(y)\}$. 
Its mean is $\mu$ and its standard deviation is $\sigma$, 
and $\ga_j=\E H_j((X-\mu)/\sigma)$. 
Note that $\ga_0=1$ and that $\ga_1=\ga_2=0$, while  
$$\ga_3=\E\Bigl({X-\mu\over \sigma}\Bigr)^3,\,\,
\ga_4=\E\Bigl({X-\mu\over \sigma}\Bigr)^4,\,\,
\ga_5=\E\Bigl({X-\mu\over \sigma}\Bigr)^5
	-10\,\E\Bigl({X-\mu\over \sigma}\Bigr)^3, $$
and so on, featuring skewness, kurtosis, pentakosis and so on, 
all of which are zero for the normal density. 
Any density with finite moments can be approximated 
with one of the form (4.6), through inclusion of enough terms. 
See Hjort \& Jones (1994) 
for details pertaining to this and some of the following calculations. 

Assume that the true $f$ is as in (4.4)
and that the normal-corrected estimator (3.5) is used, 
so that $f=f_0r$ with $f_0$ being the
simple normal approximation and $r(x)=r_0(y)$,
where $r_0(y)=\sum_{j=0}^m(\ga_j/j!)H_j(y)$, writing $y=(x-\mu)/\sigma$. 
Then 
$$\eqalign{
f''(x)&=\sigma^{-3}g''(y)=\sigma^{-3}\phi(y)\sum_{j=0}^m
	(\ga_j/j!)H_{j+2}(y), \cr 
f_0(x)r''(x)&=\sigma^{-3}\phi(y)r_0''(y) 
	=\sigma^{-3}\phi(y)\sum_{j=2}^m(\ga_j/j!)j(j-1)H_{j-2}(y). \cr}$$
Calculations give that  
$A_{j,k}=\int H_jH_k\phi^2\allowbreak\,\d y$ 
is zero when $j+k$ is odd
and equal to $(-1)^{j+p}(2\rootpi)^{-1}(2p)!/(p!2^{2p})$ 
when $j+k=2p$, see Hjort \& Jones (1994). 
This makes it possible to evaluate 
$$\eqalign{
R_{\rm trad}(f)&=\sigma^{-5}\sum_{j,k\le m}
	(\ga_j/j!)(\ga_k/k!)\,A_{j+2,k+2} \cr  
{\rm and\ }R_{\rm new}(f)&=\sigma^{-5}\sum_{2\le j,k\le m}
	{\ga_j\over (j-2)!}{\ga_k\over (k-2)!}A_{j-2,k-2} \cr} $$
for given values of $m$. 
As an example, suppose terms corresponding to 
skewness, kurtosis and pentakosis are included. Then
$$\eqalign{
R_{\rm new}
&=\sigma^{-5}\{\ga_3^2A_{1,1}+(\ga_4^2/4)A_{2,2}
	+(\ga_5^2/36)\ga_5^2+2(\ga_3\ga_5/6)A_{1,3}\} \cr
&=\sigma^{-5}{1 \over 2\rootpi}\{(1/2)\ga_3^2+(3/16)\ga_4^2
	+(5/96)\ga_5^2-(1/4)\ga_3\ga_5\} \cr
&=\sigma^{-5}{3 \over 8\rootpi}\Bigl({2\over 3}\ga_3^2
 	+{1\over 4}\ga_4^2+{5\over 72}\ga_5^2
	-{1\over 3}\ga_3\ga_5\Bigr), \cr} \eqno(4.5)$$
while Hjort \& Jones (1994) finds 
$$R_{\rm trad}
=\sigma^{-5}{3\over 8\rootpi}\Bigl(1+{35\over 48}\ga_4
	+{35\over 32}\ga_3^2
	+{385\over 1024}\ga_4^2
	+{1001\over 10240}\ga_5^2-{77\over 128}\ga_3\ga_5\Bigr).$$
This indicates that the new estimator is better than 
the traditional one for all cases in a large neighbourhood 
around the normal distribution. 

One might also use this test-bed to see where $f_0(x)r''(x)$ 
is smaller in size than $f''(x)$, say for moderate values
of $\gamma_3$, $\gamma_4$, $\gamma_5$. 
This would be analogous to the experiments described in 
Section 5A for normal mixtures.

\subsection
{\csc 4C. Second test-bed: Robust Hermite expansions.}
The Hermite expansion (4.4) is of the type 
encountered in Edgeworth--Cram\'er expansions. 
It is pleasing from a theoretic point of view 
in that it incorporates skewness, kurtosis etc.~to refine 
the normal approximation, but it has shortcomings as well.
The coefficients are not always finite, and empirical estimates
are quite variable and non-robust. 
Hjort \& Jones (1994) and Hjort \& Fenstad (1994) 
give further reasons favouring another and more robust Hermite expansion,
in terms of the polynomials $H_j^*(y)=H_j(\sqrt{2}y)$ instead.
In this case 
\def\yyy{\Bigl({x-\mu\over \sigma}\Bigr)}
$$f(x)=\phi\yyy{1\over \sigma}  
	\sum_{j=0}^m{\de_j\over j!}H_j^*\yyy, \eqno(4.6)$$
where the coefficients are determined from 
$\de_j=\sqrt{2}\E H_j(\sqrt{2}(X-\mu)/\sigma)
	\exp\{-\half(X-\mu)^2/\sigma^2\}$. 
If $f$ is taken as an approximation to a given density $q$ 
with mean $\mu$ and standard deviation $\sigma$, 
then the $L_2$ distance $\int(f-q)^2\,\d x$ is minimised 
for exactly these $\de_j$, see Hjort \& Jones (1994). 
For this expansion,  
$f_0(x)r''(x)=\sigma^{-3}\phi(y)\sum_{j=2}^m2j(j-1)(\de_j/j!)H_{j-2}^*(y)$. 
It follows from this that
$$R_{\rm new}(f)=\sigma^{-5}{2\over \rootpi}
	\sum_{j=0}^{m-2}\de_{j+2}^2/j!\,. \eqno(4.7)$$ 
An expression for $R_{\rm trad}(f)$ for this robust Hermite
expansion is in Hjort \& Jones (1994). For illustration consider the
4th order case, where terms having $\de_0,\ldots,\de_4$ are included. Then 
$$R_{\rm trad}(f)
={\sigma^{-5}\over 8\rootpi}
\Bigl(3\de_0^2+15\de_1^2+{39\over 2}\de_2^2+{25\over 2}\de_3^2
	+{41\over 8}\de_4^2-12\de_0\de_2-20\de_1\de_3-14\de_2\de_4
	+2\de_0\de_4\Bigr), $$
while $R_{\rm new}(f)=(2\sigma^{-5}/\rootpi)(\de_2^2+\de_3^2+\half\de_4^2)$.
Again this indicates superiority of the (3.5) estimator in a 
broad neighbourhood around the normal. 


\bigskip
{\bf 5. Exact analysis for normal mixtures.} 
Consider a normal mixture 
$$f(x)=\sum_{i=1}^kp_if_i(x),
\quad{\rm where\ }f_i(x)=\phi_{\sigma_i}(x-\mu_i), \eqno(5.1)$$
writing $\phi_\sigma(u)=\sigma^{-1}\phi(\sigma^{-1}u)$. 
The family of such mixtures form a very wide and flexible 
class of densities. 
Marron \& Wand (1992) studied such mixtures and in particular 
singled out 15 different `test densities',
covering a broad spectrum of not so difficult to extremely 
difficult cases, see the figure.
These will now be used by us to compare the new normal-start times
correction method with the traditional kernel method. 
In 5A the asymptotic mean squared errors of the two methods 
are compared, involving the leading terms of the Taylor-based 
approximations to bias and variance. In 5B we go further 
and analyse exact finite-sample mean squared errors for the two methods. 

\subsection
{\csc 5A. Exact amise analysis.}  
To monitor the two bias terms we should compare $f''$ to $f_0r''$,
where $f_0$ is the best approximating normal, with  
$\mu_0=\sum_{i=1}^kp_i\mu_i$ 
and 
$\sigma_0^2=\sum_{i=1}^kp_i\{\sigma_i^2+(\mu_i-\mu_0)^2\}$. 
Write $f_i=\exp(g_i)$ and $f_0=\exp(g_0)$. Then 
$r=f/f_0=\sum_{i=1}^kp_i\exp(g_i-g_0)$ and 
$r''=\sum_{i=1}^kp_i\exp(g_i-g_0)\{g_i''-g_0''+(g_i'-g_0')^2\}$. 
This leads to 
$$f_0(x)r''(x)=\sum_{i=1}^kp_if_i(x)[1/\sigma_0^2-1/\sigma_i^2
	+\{(x-\mu_i)/\sigma_i^2-(x-\mu_0)/\sigma_0^2\}^2], \eqno(5.2)$$
while 
$$f''(x)=\sum_{i=1}^kp_i\phi_{\sigma_i}''(x-\mu_i)
	=\sum_{i=1}^kp_i\{(x-\mu_i)^2/\sigma_i^2-1\}
	f_i(x)/\sigma_i^2. \eqno(5.3)$$
With some efforts (5.2) and (5.3) also lead to formulae 
for the roughness values 
$R_{\rm trad}(f)$ and $R_{\rm new}(f)$, cf.~(4.2). 
Exact expressions are given in Proposition A.1 in our Appendix I. 

In the figure these formulae are used to visually inspect 
$f''(x)$ versus $f_0(x)r''(x)$, for each of the 15 test cases. 
There are two immediate points to note.
The first is that in most cases where the initial 
normal approximation is not very unreasonable, 
the new estimator manages to be better than the usual one, 
in significant $x$-areas. The second observation is that 
in cases where the initial description is clearly a bad start,
the new semiparametric method turns almost nonparametric and 
behaves almost like the kernel method. 
\fermat{For simplicity the figure
is placed at the end of our report}

{{\medskip\narrower\noindent\baselineskip11pt\sl
{\csc Figure.}
The 15 test densities 
(left hand side) presented together with the bias factor 
functions $f''$ (solid line, for the kernel method) and $f_0r''$ 
(dotted line, for the new method). 
\medskip}}

We have also computed the global criteria 
$R_{\rm trad}(f)$ and $R_{\rm new}(f)$,
for each of the 15 test densities;
see the formulae and Table A.1 in Appendix I.
The overall comparison in terms of approximate mise
is in clear favour of the new method.
Roughly speaking the first nine test cases are the not drastically 
unreasonable ones, whereas cases 10--15 probably originate 
from another planet and were chosen by Marron \& Wand 
to exhibit particularities of smoothing parameter problems. 
And the new method wins in each of the nine worldly cases:   
the Gau\ss ian, 
the skewed unimodal, 
the strongly skewed,
the kurtotic unimodal, 
the outlier,
the bimodal, 
the separated bimodal, 
the skewed bimodal, 
the trimodal. 
It is also better for the claw density (\#10 in Marron \& Wand),
the double claw (\#11), and even for 
the asymmetric double claw (\#13). It only loses 
to the traditional kernel method, and then only very slightly, in cases 
\#12 (the asymmetric claw), 
\#14 (the smooth comb), 
and \#15 (the discrete comb). 

So in terms of approximate mise the semiparametric (3.5) estimator 
wins over the kernel method in 12 out of 15 cases. It is fair to add
that only about half of these victories are clear-cut, and that 
the remaining cases are almost draws, with surprisingly 
similar values for $R_{\rm new}$ and $R_{\rm trad}$. 
This picture emerges also when one computes values for 
the $L_1$-based criteria $\int |f''|$ versus $\int |f_0r''|$,
also given in the table of Appendix I. 
According to this measure 
the (3.5) estimator wins in 14 out of 15 cases. 

We also inspected separately the case of two components in the normal
mixture. Only in quite extreme cases does the kernel method win
in approximate mise, and then only slightly. 
And the new method always wins when the two standard
deviation parameters in question are equal. 
It is mildly surprising that a nonparametric correction on 
a normal start performs better than the kernel method even in 
such highly non-normal situations. 

\subsection
{\csc 5B. Exact finite-sample comparison.}
The comparison analysis above was in terms 
of the Taylor-based approximations to bias and variance. 
Now we go further and analyse exact finite-sample mise 
for the two methods. 
Such analysis was carried out in Marron \& Wand (1992) for 
the kernel method (1.1). 
Their Theorem 2.1 implies that if $f$ is as in (5.1), then 
$$\eqalign{{\rm mise}(h)
&=\E\int(\tilda f-f)^2\,\d x \cr
&=\Bigl(1-{1\over n}\Bigr)\sum_{i,j}
{p_ip_j\over (\sigma_i^2+\sigma_j^2+2h^2)^{1/2}}
\phi\Bigl({\mu_j-\mu_i\over (\sigma_i^2+\sigma_j^2+2h^2)^{1/2}}\Bigr) \cr
&\qquad\qquad
+{1\over n}{1\over 2\rootpi h} 
-2\sum_{i,j}{p_ip_j\over (\sigma_i^2+\sigma_j^2+h^2)^{1/2}}
\phi\Bigl({\mu_j-\mu_i\over (\sigma_i^2+\sigma_j^2+h^2)^{1/2}}\Bigr) \cr
&\qquad\qquad 
+\sum_{i,j}{p_ip_j\over (\sigma_i^2+\sigma_j^2)^{1/2}}
\phi\Bigl({\mu_j-\mu_i\over 
	(\sigma_i^2+\sigma_j^2)^{1/2}}\Bigr). \cr}\eqno(5.4)$$
Reaching a similar result for the mise of the normal-start estimator 
(3.5) is much more demanding. 
Proposition A.2 in Appendix II delivers such a formula. 
It simplifies the comparison quest to care only about 
`best case versus best case',
which means comparing the two best achievable mise values, 
say ${\rm mise}_{\rm trad}^*$ and ${\rm mise}^*$. 
We programmed formula (5.4) and the one in Proposition A.2 
and went through the list of the 15 test densities again, 
and found for each the minimising value of $h$ 
and the resulting minimum mise values, 
for each of the five sample sizes 25, 50, 100, 200, 1000. 
The results are displayed in Table A.2 of Appendix II,
along with the ratio ${\rm mise}^*/{\rm mise}^*_{\rm trad}$.  
These numbers support the previous positive conclusions for 
the new estimator, in its particular form (3.5). 
The mise-ratio is quite often below 1, and for the 
quite difficult test densities, where the analysis of 5A 
gave very similar values for $R_{\rm trad}$ and $R_{\rm new}$, 
Table A.2 yields mise-ratios mostly between 0.99 and 1.01.
Even in these highly non-normal situations the new method 
has, overall, a slight edge. 
The table also illustrates that choosing the same bandwidth 
for the new method as for the kernel method will be quite
acceptable in most of the definitely non-normal situations. 
In a broad vicinity of the normal it should pay to use 
a little larger bandwidth than what is optimal for the kernel method, however. 

It should be kept in mind that the list of 15 test densities 
is not at all constructed to be favourable to using the 
normal model as start description. Statistically speaking 
we believe that a high proportion of densities actually 
encountered in real life are closer to the normal than 
each of cases \#3--\#15. In other words, the new method will 
win quite often. 
 
\bigskip
{\bf 6. Choosing smoothing parameter.}
Our method is defined in terms of a kernel function $K$
and a bandwidth or smoothing parameter $h$. 
Choosing $h$ is the more crucial problem, 
and methods for doing this parallel but by necessity become harder 
than the well-developed ones for the traditional (1.1) 
estimator (which is the special case of a constant initial estimator).  

\subsection
{\csc 6A. Minimising amise.}
From (4.1) it is seen that the $h$ parameter minimising 
approximate integrated mean squared error for $\hatt f$ is 
$$h=h^*=\{R(K)/\sigma_K^4\}^{1/5}
	R_{\rm new}(f)^{-1/5}\,n^{-1/5}. \eqno(6.1)$$
The resulting minimal amise is 
${5\over 4}\{\sigma_KR(K)\}^{4/5}R_{\rm new}^{1/5}n^{-4/5}$. 
The same $\{\sigma_KR(K)\}^{4/5}$ factor appears also in 
a similar expression for the theoretically best point-wise
mean squared error, so the efficiency of the kernel choice lies 
entirely with this number. This is very similar to what happens
with the traditional estimator (1.1), see Scott (1992, Chapter 6),
for example. 
The best possible kernel in this sense is the Yepanechnikov 
kernel $K_0(z)={3\over 2}(1-4z^2)$ supported on 
$[-\half,\half]$ (or any other scaled version). 

A `plug-in rule' for $h$ is to estimate the roughness 
$R_{\rm new}$ of (4.2) and insert this into (6.1). 
We outline three methods for doing this. 

The first method is in the parametric `rule of thumb' tradition 
and fits the data initially to a normal mixture, say of two or three 
components, using likelihood-based methods. The idea is then to
use the formula for $R_{\rm new}$ in Appendix I to estimate $h^*$ of (6.1). 
This would work well in many cases. 

The second method is to exploit the Hermite expansions 
of Section 4 as approximations to the true $f$. 
An approximation to $f$ that takes the first five moments
into account is (4.4), with empirical estimates 
inserted for $\ga_3$, $\ga_4$, $\ga_5$. This leads to an estimate
of $R_{\rm new}$ via (4.5). The result, in the case of 
the normal kernel $K=\phi$, becomes 
$$\hatt h_1=(4/3)^{1/5}\{(2/3)\hatt\ga_3^2+(1/4)\hatt\ga_4^2
	+(5/72)\hatt\ga_5^2-(1/3)\hatt\ga_3\hatt\ga_5\}^{-1/5}
	\hatt\sigma\,n^{-1/5}. \eqno(6.2)$$
One should preferably use robust estimates for the parameters,
and one should ideally also deduct for bias when plugging in 
squared estimates, as explained in Hjort \& Jones (1994). 
In any case (6.2) may be somewhat unstable, particularly
for small to moderate sample sizes, since the 
empirical $\hatt\ga_j$ statistics are unstable. 
The alternative robust Hermite expansion described in 4D 
should be safer, using (4.6)--(4.7) instead of (4.4)--(4.5). 
It uses the automatically robust estimates
$$\hatt\de_j={1\over n}\sumin \sqrt{2}H_j
	\Bigl(\sqrt{2}{X_i-\hatt\mu\over \hatt\sigma}\Bigr)
	\exp\Bigl\{-\half\Bigl({X_i-\hatt\mu\over \hatt\sigma}\Bigr)^2\Bigr\}$$
(the summands are bounded in $X_i$) and 
$$\hatt h_2=(1/4)^{1/5}
(\hatt\de_2^2+\hatt\de_3^2+\hatt\de_4^2/2+\hatt\de_5^2/6)^{-1/5}
	\hatt\sigma\,n^{-1/5}, \eqno(6.3)$$
for example. Again bias should ideally be deducted when plugging in squared 
estimates. See analogous comments in Hjort \& Jones (1994). 

While this second method can be seen as a semiparametric
way of getting hold of $R_{\rm new}$, the 
third plug-in method is nonparametric on this account and 
takes the natural statistic
$$\eqalign{
\hatt R_{\rm new}&=\int \{f(x,\hatt\theta)\hatt r''(x)\}^2\,\d x \cr 
&={1\over n^2}{1\over h^6}\sum_{i,j}\int
{f(x,\hatt\theta)\over f(X_i,\hatt\theta)}
{f(x,\hatt\theta)\over f(X_j,\hatt\theta)}
	K''(h^{-1}(x-X_i))K''(h^{-1}(x-X_j))\,\d x \cr}$$
as its starting point. 
Explicit expressions for the integral here
can be worked out for most choices of $K$;
see formula (A.6) in Appendix II.   
Using (3.2) and the techniques of Section 3 one can show that 
$$\eqalign{
R^*_{\rm new}&=\int \{f_0(x)(r^*)''(x)\}^2\,\d x \cr 
&={1\over n^2}{1\over h^6}\sum_{i,j}\int
{f_0(x)\over f_0(X_i)}{f_0(x)\over f_0(X_j)}
	K''(h^{-1}(x-X_i))K''(h^{-1}(x-X_j))\,\d x, \cr}$$
in which $f_0(x)=f(x,\theta_0)$ and 
$r^*(x)=n^{-1}\sumin K_h(X_i-x)/f_0(X_i)$, 
is a good approximation to $\hatt R_{\rm new}$;
in particular the mean of $\hatt R_{\rm new}$ is only $O(h^2/n+n^{-2})$
away from the mean of $R^*_{\rm new}$. 
Now somewhat long calculations, involving Taylor expansions,
can be furnished to reach 
$$\E R^*_{\rm new}={n-1\over n}\int (f_0r'')^2\,\d x
	+{1\over nh^5}\{R(K'')+O(h^2)\}, $$
where $R(K'')=\int(K'')^2\,\d z$. Since $nh^5$ is stable 
this shows that there is a fixed amount of overshooting. 
This is similar to but more involved than the corresponding
result for the traditional kernel estimator (1.1)
(which is the special case where $f_0(x)$ is constant),
see Scott \& Terrell (1987). This invites 
${n\over n-1}\{\hatt R_{\rm new}-R(K'')/(nh^5)\}$ to be used 
as a corrected estimate. One version of the plug-in method
is therefore as follows: Select a start value for $h$   
in a reasonable way, perhaps using (6.3). Then compute 
$\hatt R_{\rm new}$ and its de-biased version, and insert in (6.1).
One might also iterate this scheme further. 

It is required that $K$ here is smooth with vanishing derivatives 
at the end points of its support; in particular the Yepanechnikov 
kernel is not allowed in this operation. 


\subsection
{\csc 6B. Minimising estimated amise.} 
A useful idea related to the previous calculations is to
estimate the approximate mise of (4.1) directly, that is, 
producing the curve 
$$\hatt{\rm amise}(h)={\rm bcv}(h)
	=\quart\sigma_K^4h^4\Bigl\{\hatt R_{\rm new}(h)
	-{R(K'')\over nh^5}\Bigr\}
	+{R(K)\over nh}, \eqno(6.4)$$
including for emphasis $h$ in the notation for the roughness estimate. 
This function must now be computed for a range of $h$-values,
up to some upper limit $h_{\rm os}$, the `over-smoothing' bandwidth.   
Scott \& Terrell (1987) and Scott (1992) call this strategy 
(for the traditional estimator) 
`biased cross validation', although nothing seems to be 
cross validated per se. The bcv name derives rather from 
formula-wise similarity to unbiased cross validation, see below, 
and the desire to estimate the biased approximation amise to 
the true mise. 

\subsection
{\csc 6C. Nearly unbiased cross validation.}
A popular technique for the traditional kernel estimator 
is that of unbiased least squares cross validation,
minimising an unbiased estimate of the exact mise 
as a function of bandwidth. A version of this idea can
be carried through for our new estimator as well. 
The crux is to estimate  
${\rm mise}(h)-R(f)=\E\{\int\hatt f^2\,\d x-2\int f\hatt f\,\d x\}$ 
with 
$${\rm ucv}(h)=\int\hatt f_h(x)^2\,\d x
	-{2\over n}\sumin \hatt f_{h,(i)}(X_i). \eqno(6.5)$$
Here $h$ is included in the notation for clarity, and 
$\hatt f_{h,(i)}$ is the estimator constructed from 
the diminished data set that excludes $X_i$. The function to compute is 
$$\eqalign{
{\rm ucv}(h)&={1\over n^2}\sum_{i,j}
{1\over f(X_i,\hatt\theta)f(X_j,\hatt\theta)}
\int f(x,\hatt\theta)^2K_h(x-X_i)K_h(x-X_j)\,\d x \cr
&\qquad\qquad 
-{2\over n(n-1)}\sum_{i,j}K_h(X_i-X_j)
	{f(X_i,\hatt\theta_{(i)})\over f(X_j,\hatt\theta_{(j)})}, \cr}$$
where $\hatt\theta_{(i)}$ is computed without $X_i$. 
In the case of the normal start method (3.5) with normal kernel $K=\phi$
a formula for the first term here is given in (A.6) in the Appendix.

It turns out that ${\rm ucv}(h)$ is nearly but not exactly unbiased 
for ${\rm mise}(h)-R(f)$. We have 
$$\eqalign{
\E\int f\hatt f\,\d x
&=\E\int f(x)K_h(X_1-x)
	{f(x,\hatt\theta)\over f(X_1,\hatt\theta)}\,\d x \cr
&=\E\int\int f(x)f(y)K_h(y-x){f(x,\hatt\theta(y,X_2,\ldots,X_n))
	\over f(y,\hatt\theta(y,X_2,\ldots,X_n))}\,\d x\,\d y, \cr}$$
which is subtly different from 
$$\eqalign{
\E{1\over n}\sumin\hatt f_{(i)}(X_i)
&=\E K_h(X_2-X_1){f(X_1,\hatt\theta(X_2,\ldots,X_n))
	\over f(X_2,\hatt\theta(X_2,\ldots,X_n))} \cr
&=\E\int\int f(x)f(y)K_h(y-x){f(x,\hatt\theta(y,X_3,\ldots,X_n))
	\over f(y,\hatt\theta(y,X_3,\ldots,X_n))}\,\d x\,\d y. \cr}$$ 
The difference is minuscule, however, and choosing $h$ to minimise
the ${\rm ucv}(h)$ function, among $h\le h_{\rm os}$ for a
suitable over-smoothing upper limit, remains a useful 
and honestly nonparametric option. 

\subsection
{\csc 6D. Other techniques.}
Other techniques can also be proposed, for example trying to 
adapt recent methods of Sheather \& Jones (1991) and of 
Hall, Sheather, Jones \& Marron (1991) to the present 
situation. One could also look into possible advantages of 
using a variable $h$. These matters are not pursued here. 
In our somewhat limited experience the (6.3) method has been 
satisfactory. 

\def\bfX{{\bf X}}
\def\bfx{{\bf x}}
\def\bfh{{\bf h}}
\def\bfY{{\bf Y}}
\def\bfy{{\bf y}}

\bigskip
{\bf 7. The multi-dimensional case.} 
Our multiplicative correction factor method works well 
also in the vector case, as is now briefly explained. 
The setting is that $d$-dimensional
i.i.d.~vectors $\bfX_1,\ldots,\bfX_n$ are observed from 
a density $f$. 

\subsection
{\csc 7A. The traditional and the new estimator.} 
The traditional kernel estimator 
uses a kernel density function $K(z_1,\ldots,z_d)$, 
usually symmetric about zero in each direction and often
of product form $K_1(z_1)\cdots K_d(z_d)$. Its value 
at the point $\bfx=(x_1,\ldots,x_d)'$ is 
$$\tilda f({\bf x})={1\over n}\sumin K_{\bfh}
	(X_{i,1}-x_1,\ldots,X_{i,d}-x_d), \eqno(7.1)$$
where $K_{\bfh}(z_1,\ldots,z_d)=(h_1\cdots h_d)^{-1}
K(h_1^{-1}z_1,\ldots,h_d^{-1}z_d)$; 
see for example Scott (1992, Chapter 6) 
or Wand \& Jones (1994, Chapter xx). 
In the product kernel case 
the basic bias and variance behaviour is described by 
$$\eqalign{
{\rm bias}&\doteq \half\sum_{j=1}^d\sigma(K_j)^2h_j^2f''_{jj}(\bfx), \cr 
{\rm variance}&\doteq R(K_1)\cdots R(K_d)(nh_1\cdots h_d)^{-1}f(\bfx)
	-n^{-1}f(\bf x)^2, \cr} \eqno(7.2)$$
where $\sigma(K_j)^2=\int z^2K_j(z)\,\d z$ and $R(K_j)=\int K_j(z)^2\,\d z$.
Furthermore $f''_{jj}$ is the second partial derivative of $f$ 
in direction $x_j$. 

Our parametric start with a multiplicative correction method is now 
$$\hatt f(\bfx)=f(\bfx,\hatt\theta){1\over n}\sumin 
	K_{\bfh}(\bfX_i-\bfx)/f(\bfX_i,\hatt\theta). \eqno(7.3)$$
This is the appropriate vector version of (1.3), 
employing any parametric family $f(\bfx,\theta)$ 
and any reasonable parameter estimation method 
to produce the initial $f(\bfx,\hatt\theta)$. 
The most important case is that of 
a multinormal start density, in which case the new estimator is 
$$\hatt f(\bfx)={1\over n}\sumin 
K_{\bfh}(\bfX_i-\bfx)
	{\exp\{-\half(\bfx-\hatt\mu)'\hatt\Sigma^{-1}(\bfx-\hatt\mu)\}
	\over \exp\{-\half(\bfX_i-\hatt\mu)'\hatt\Sigma^{-1}
	(\bfX_i-\hatt\mu)\}}, \eqno(7.4)$$
and with some computational simplifications possible if 
a Gau\ss ian kernel is used. 

One may now go through the theory developed in Sections 2 and 3 
and generalise results there to the present $d$-dimensional state
of affairs. We omit details and merely present the result. 
Firstly, the variance of the (7.2) estimator is found to be 
exactly equal to the variance noted above
for the traditional (7.1) estimator, to the order of approximation used. 
Secondly, the bias is of the form 
$$\half\sum_{j=1}^d\sigma(K_j)^2h_j^2f_0(\bfx)r''_{jj}(\bfx)
	+O\bigl(\sum_{j=1}^d(h_j^4+h_j^2/n)+n^{-2}\bigr), $$
involving the best parametric approximant $f_0(\bfx)=f(\bfx,\theta_0)$
and the ensuing correction factor $r(\bfx)=f(\bfx)/f_0(\bfx)$. 
Again the result is remarkably resistant to the actual parameter
estimation used to obtain $\hatt\theta$, for example, cf.~the discussion
of Section 3. 

Method (7.3) can therefore be expected 
to perform well in all situations where the $f_0r''_{jj}$ functions
are smaller in size than the $f''_{jj}$ functions. 
This essentially says that the correction factor $r$ should 
have smaller sized curvature than $f$ itself, which again
means that the initial parametric description should 
capture the main features of the density.   
Special cases can be inspected as explained in Sections 4 and 5. 
We expect the attractive (7.4) method, for example, 
in which case $f_0$ becomes the multinormal with parameters equal to 
the true mean and true covariance matrix for $f$,
to work better than the traditional (7.1) estimator, 
for densities in a broad nonparametric vicinity of the multinormal. 

\def\quotationone{\smallrm{Il Cervo Gaio}}
\def\quotationtwo{\smallrm Semiparametric density estimation} 

\subsection 
{\csc 7B. A particular scheme.} 
We speculate that the new methods could prove to be particularly useful 
in higher dimensions, since the traditional estimators, like (7.1),
have quite slow convergence rates then. Implementation of the (7.4)
estimator is straightforward, but the smoothing parameters 
remain to be specified. This is a harder problem than in the
one-dimensional case. For completeness we briefly describe one 
particular solution here. It is practical and should work well
in many situations, but does not claim optimality. 

Start out considering the density $g(\bfy)$ of 
$\bfY_i=\Sigma^{-1/2}(\bfX_i-\mu)$,
where $\mu$ and $\Sigma$ are mean vector and covariance matrix for $\bfX_i$. 
Since these `sphered' variables have mean zero and covariance matrix 
the identity a natural start description of $g$ is $g_0$, the
standard multi-normal, and furthermore  
it appears reasonable to smooth with the same amount in each
direction, for example using the standard multi-normal 
$K_h({\bf z})=h^{-d}(2\pi)^{-d/2}\exp(-\half\|{\bf z}\|^2/h^2)$. 
An estimate of $g$ would consequently be of the form 
$\hatt g(\bfy)=g_0(\bfy)\,n^{-1}\sumin K_h({\bfY}_i-\bfy)/g_0({\bfY}_i)$. 
After estimating mean and covariance matrix 
this amounts to an estimated multinormal start for $f$ and leads to 
$$\eqalign{\hatt f(\bfx)
&=\hatt g(\hatt\Sigma^{-1/2}(\bfx-\hatt\mu))|\hatt\Sigma|^{-1/2} \cr
&={1\over n}\sumin {\exp\{-\half(\bfx-\bfx_i)'\hatt\Sigma^{-1}
	(\bfx-\bfx_i)/h^2\}\over (2\pi)^{d/2}h^d}
  {\exp\{-\half(\bfx-\hatt\mu)'\hatt\Sigma^{-1}(\bfx-\hatt\mu)\}
  \over 
  \exp\{-\half(\bfX_i-\hatt\mu)'\hatt\Sigma^{-1}(\bfX_i-\hatt\mu)\}}. 
	\cr} \eqno(7.5)$$
This estimator can also be motivated directly without $\bfY_i$s.
The main reason for using the $g$-representation is however 
that it can be used to find a suitable $h$, as follows.  
By previous results the approximate mise is 
$R(\phi)^d(nh^d)^{-1}+\quart h^4R_{\rm new}(g)$, featuring  
$R_{\rm new}(g)=\int \{g_0(\bfy)r''(\bfy)\}^2\,\d\bfy$. 
Its minimiser 
$h^*=\{dR(\phi)^d\}^{1/(d+4)}R_{\rm new}(g)^{-1/(d+4)}
	\allowbreak\,n^{-1/(d+4)}$ 
can be estimated in various ways,
and one feasible solution, aiming to generalise (4.7) and (6.3), 
is to approximate $r$ using an
expansion with products $H_{j_1}^*(y_1)\cdots H_{j_d}^*(y_d)$ 
as basis functions, where again $H_j^*(y)=H_j(\sqrt{2}y)$. 
We omit the many necessary details here but report that 
$$R_{\rm new}(g)={4\over (2\rootpi)^d}\sum_{j_1,\ldots,j_d}
	(\de_{j_1+2,\ldots,j_d}+\cdots+\de_{j_1,\dots,j_d+2})^2
	/(j_1!\cdots j_d!), $$
where 
$$\de_{j_1,\ldots,j_d}=2^{d/2}\E_g\exp(-\half\|\bfY\|^2)\,
	H_{j_1}^*(Y_1)\cdots H_{j_d}^*(Y_d). $$	
The procedure is as with (4.7) and (6.3), namely estimating 
the first few of these using 
$$\hatt\de_{j_1,\ldots,j_d}=2^{d/2}{1\over n}\sumin 
	\exp\{-\half(\bfX_i-\hatt\mu)'\hatt\Sigma^{-1}(\bfX_i-\hatt\mu)\}\,
	H_{j_1}^*(Y_{i,1})\cdots H_{j_d}^*(Y_{i,d}), $$	
where now $\bfY_i=\hatt\Sigma^{-1/2}(\bfX_i-\hatt\mu)$, and finally 
calculating 
$$\eqalign{\hatt h
&=\{d/(2\rootpi)^d\}^{1/(d+4)}\hatt R_{\rm new}^{-1/(d+4)}n^{-1/(d+4)} \cr
&=(d/4)^{1/(d+4)}\Bigl\{\sum_{j_1,\ldots,j_d}
	{(\hatt\de_{j_1+2,\ldots,j_d}+\cdots+\hatt\de_{j_1,\dots,j_d+2})^2
	\over j_1!\cdots j_d!}\Bigr\}^{-1/(d+4)}\,n^{-1/(d+4)}. \cr}$$


\bigskip
{\bf 8. Supplementing remarks.}

\subsection
{\csc 8A. How close is the new estimator to the old?} 
For simplicity of presentation consider $\hatt f(x)$ in the form
(2.1) in terms of a basis function $f_0(x)$ 
rather than with estimated parameters. In the sum that defines 
$\hatt f(x)$ the ratios $f_0(x)/f_0(X_i)$ 
are close to 1 for small values of $h$ since then the $X_i$s 
quite close to $x$ are those given significant weights.
In other words, $\hatt f(x)$ cannot be very different from 
the traditional kernel estimator $\tilda f(x)$ of (1.1) when $h$ is small. 
A Taylor analysis is informative:
$${f_0(x)\over f_0(X_i)}\doteq 1-a_0(x)(X_i-x)
	+\half\{a_0(x)^2-b_0(x)\}(X_i-x)^2, $$
where $a_0(x)$ and $b_0(x)$ are the two first $x$-derivatives 
of $\log f_0(x)$. Hence 
$$\hatt f(x)\doteq \tilda f(x)-a_0(x)e_1(x)
	+\half\{a_0(x)^2-b_0(x)\}e_2(x), \eqno(8.1)$$
where $e_q(x)=n^{-1}\sumin K_h(X_i-x)(X_i-x)^q$. 
One can now show that $e_1(x)$ has mean $\sigma_K^2h^2f'(x)+O(h^4)$ 
and small variance $O(hn^{-1}f(x))$, while 
$e_2(x)$ has mean $\sigma_K^2h^2f(x)+O(h^4)$ with even smaller 
variance $O(h^3n^{-1}f(x))$. Thus the difference is of size $O(h^2)$. 
If $f_0(x)$ is the standard normal, for example, then 
$\hatt f(x)-\tilda f(x)=h^2\sigma_K^2\{xf'(x)+\half(x^2+1)f(x)\}+O_p(h^4)$. 
 

\subsection
{\csc 8B. Accuracy of the estimated correction factor.}
Our machinery can also be used for model exploration purposes,
by inspecting the correction factor against $x$ 
for various potential models. A model's adequacy could
be inspected by looking at a plot of $\hatt r(x)$,
perhaps with a pointwise confidence band, to see if $r(x)=1$ 
is reasonable. 
In the notation of Sections 2 and 3, and using techniques from these
sections, one can establish that 
$$\eqalign{
\E\,\hatt r(x)&\doteq r(x)+\half\sigma_K^2h^2r''(x)
	-n^{-1}r(x)u_0(x)'\{I(x)+d\}, \cr
\Var\,\hatt r(x)&\doteq (nh)^{-1}R(K)r(x)/f_0(x)-n^{-1}r(x)^2
	\{1+2u_0(x)'I(x)-u_0(x)'\Sigma u_0(x)\}, \cr}$$
with some simplification in the maximum likelihood case,
for which $I(x)=J^{-1}u_0(x)$ and $\Sigma=J^{-1}$. 

It is also informative to plot the log-correction factor
$\log\hatt r(x)$, to see how far from zero it is. The bias and variance
results for this curve are
$$\eqalign{
\E\log\hatt r(x)&\doteq \log r(x)+\half\sigma_K^2h^2r''(x)/r(x) \cr
	&\qquad\qquad
	-\half R(K)(nh)^{-1}\{r(x)f_0(x)\}^{-1}
	-\hbox{$1\over8$}\sigma_K^4h^4r''(x)^2/r(x)^2, \cr
\Var\log\hatt r(x)&\doteq R(K)(nh)^{-1}\{r(x)f_0(x)\}^{-1}
	-n^{-1}\{1+2u_0(x)'I(x)-u_0(x)'\Sigma u_0(x)\}. \cr}$$
A nice graphical goodness of fit method emerges: plot  
$$Z(x)={\log\hatt r(x)+\half R(K)(nh)^{-1}f(x,\hatt\theta)^{-1}
\over \{R(K)(nh)^{-1}f(x,\hatt\theta)^{-1}\}^{1/2}} \eqno(8.2)$$
against $x$, possibly with a more accurate denominator.  
Under model conditions this should be approximately 
distributed as a standard normal for each $x$, that is,
the $Z(x)$ curve should stay within $\pm1.96$ about 95\% of the time. 


\def\quotationone{\smallrm{Glad Hjort}}
\def\quotationtwo{\smallrm Semiparametric density estimation} 

\subsection
{\csc 8C. The integral.}  
Our estimator does not integrate to precisely 1. 
The normal-based version (3.5), for example, 
when the Gau\ss ian kernel $K=\phi$ is used, has  
$$\int\hatt f\,\d x=(1+h^2/\hatt\sigma^2)^{-1/2}
{1\over n}\sumin\exp\{\half h^2(X_i-\hatt\mu)^2/
	\{\hatt\sigma^2(\hatt\sigma^2+h^2)\}\}, $$
which after Taylor expansions is found to be equal to 
$1+{1\over 8}\hatt\ga_4h^4/\hatt\sigma^4$, 
where
$$ \hatt\ga_4=n^{-1}\sumin\{(X_i-\hatt\mu)/\hatt\sigma\}^4-3 $$
is the estimated kurtosis. Dividing the original estimate
with this amount does not lead to superior performance
in terms of mise, however. 
In the general case, in the notation of (2.1), for example, one finds  
$$\int \hatt f\,\d x=1+\half h^2\sigma_K^2{1\over n}
	\sumin{f_0''(X_i)\over f_0(X_i)}
+\twentyfourth h^4\E_KZ^4{1\over n}\sumin{f_0^{(4)}(X_i)\over f_0(X_i)} $$
via Taylor expansions. The $h^2$ term vanishes in the normal case. 

\subsection
{\csc 8D. Parametric home-turf conditions.}
If model conditions $f(x)=f(x,\theta)$ can be trusted 
the natural estimator is simply $f(x,\hatt\theta)$, 
for example with the maximum likelihood estimator. 
From $\sqrt{n}(\hatt\theta-\theta)\arr_d\normal_p\{0,J^{-1}\}$
where $J$ is the information matrix, 
combined with the delta method and some extra arguments, follows 
$$n\E\int\{f(x,\hatt\theta)-f(x,\theta)\}^2\,\d x
	\arr_d\int f(x,\theta)^2u(x,\theta)'J^{-1}u(x,\theta)\,\d x, $$
where $u(x,\theta)=\dell\log f(x,\theta)/\dell\theta$ is the score
function. Algebraic calculations for the normal model 
lead to a parametric mise of size ${7\over8}(2\rootpi\sigma)^{-1}/n$. 
(This is the large-sample approximation to the exact mise, 
for which an exact formula also can be found.) 

It turns out that the mise of the new nonparametric (3.5) estimator,
computed under Gau\ss ian home turf conditions, 
is only slightly larger than this. Going through formulae (A.8)--(A.10)
of Appendix II one finds 
$${\rm mise}(h)={1\over 2\rootpi}\Bigl\{\Bigl(1-{1\over n}\Bigr)
	{1\over \sigma}+{1\over n}{1\over h}{\sigma\over (\sigma^2-h^2)^{1/2}}
	-{2\over \sigma}\Bigr\}. $$
It is of separate interest to note that this is minimised for 
$h^*=\sigma/\sqrt{2}$, regardless of sample size; 
cf.~Case \#1 in Table A.2. The minimum value is 
${\rm mise}^*=(2\rootpi\sigma)^{-1}/n$, only 14\% larger than
the parametric mise. Of course one should use an estimated 
$\hatt\sigma/\sqrt{2}$ in practice, but this can be seen
to alter the minimum mise only to second order terms $O(n^{-2})$.
This is shown via an exact formula for 
${\rm ise}(\hatt\sigma/\sqrt{2})$, using results appearing after 
(A.6) in Appendix~II. 

\subsection
{\csc 8E. Nonparametric regression with a parametric start.}
The basic estimation idea of our paper works well also in other
areas of curve smoothing. An important such area is that of 
nonparametric regression. Assume that i.i.d.~pairs $(x_i,y_i)$ are observed
from a smooth bivariate density $f(x,y)=f(x)g(y\midd x)$, 
and that interest focuses on the conditional mean function
$m(x)=\E(Y\midd x)$. A standard method is the Nadaraya--Watson estimator 
$\tilda m(x)=\sumin y_iK_h(x-x_i)/\sumin K_h(x-x_i)$, 
see for example Scott (1992, Chapter 8) and Wand \& Jones (1994, Chapter xx). 
Taylor expansion analysis and somewhat lengthy calculations lead to 
$$\eqalign{
\E\,\tilda m(x)&\doteq m(x)+\half\sigma_K^2h^2\{m''(x)+2m'(x)f'(x)/f(x)\}, \cr
\Var\,\tilda m(x)&\doteq R(K)(nh)^{-1}\sigma(x)^2/f(x)
	+O(h/n). \cr}\eqno(8.3) $$
This is a somewhat more complete version of calculations 
in Scott (1992, p.~223--224). Our calculations are also mildly more general, 
in that we took care here not to assume merely a constant
value for $\sigma(x)^2=\Var(Y\midd x)$, for reasons appearing below.

A semiparametric estimator can now be constructed as follows.
Start out with a parametric initial description, say $m(x,\hatt\beta)$,
perhaps the simple linear $\hatt\beta_1+\hatt\beta_2x$. 
This start estimator aims really at $m(x,\beta_0)$, say, 
the best parametric approximant. 
A multiplicative correction factor, aiming at $r(x)=m(x)/m(x,\beta_0)$, 
can be given as a Nadaraya--Watson estimator using $y_i/m(x_i,\hatt\beta)$. 
This leads to a generalised Nadaraya--Watson estimator 
$$\eqalign{
\hatt m(x)&=m(x,\hatt\beta)
{\sumin \{y_i/m(x_i,\hatt\beta)\}K_h(x-x_i)
	\over \sumin K_h(x-x_i)} \cr
&={\sumin y_i\,\{m(x,\hatt\beta)/m(x_i,\hatt\beta)\}K_h(x-x_i)
	\over \sumin K_h(x-x_i)}. \cr}\eqno(8.4)$$
Calculations involving the above result, using $y_i/m(x_i,\beta_0)$ 
with conditional variance $\sigma(x_i)^2\allowbreak/ m(x_i,\beta_0)^2$, and
Taylor expansions of $\hatt\beta$ around $\beta_0$, 
as in Section 3, lead in the end to 
$$\E\,\hatt m(x)\doteq m(x)+\half\sigma_K^2h^2\{m(x,\beta_0)r''(x)^2
	+2m_0(x,\beta)r'(x)f'(x)/f(x)\}, \eqno(8.5)$$
with approximation error of size at most $O(h^4+h^2/n+n^{-2})$,  
and to a variance being of the very same size as that in (8.3),
to the order of approximation used. 
In many cases this will mean a genuine reduction of mise, and hence that 
the generalised Nadaraya--Watson estimator (8.4) 
is better than the usual estimator. 
This idea could be particularly useful in situations with 
several covariates.  

Hjort (1993, final section) gives yet another example of 
the type (3.1) construction, in the realm of nonparametric 
hazard rate estimation. The result is once again that 
a bias reduction vis-\`a-vis the traditional estimator 
is possible in a broad neighbourhood of the parametric model
used, without sacrificing variance. 

%

\bigskip
\bigskip
{\bf Appendix I: Roughness measures for normal mixtures.} 
Here formulae are provided for roughnesses $R_{\rm trad}$ and
$R_{\rm new}$ for a general normal mixture, results that were
used in Section 4B. A table comparing the performance of 
the normal-start semiparametric method with that of 
the usual kernel estimator, for each of 15 test cases, 
is also given. 

{{\smallskip\sl
{\csc Proposition A.1.} 
For a normal mixture $f(x)=\sum_{i=1}^k p_i\phi_{\sigma_i}(x-\mu_i)$,
let $\sigma_{i,j}^2=\sigma_i^2+\sigma_j^2$ and 
$\delta_{i,j}=(\mu_j-\mu_i)/\sigma_{i,j}$. 
The roughness functionals defined in (4.2) can be calculated explicitly;
$$\eqalign{
R_{\rm trad}&=\int(f'')^2\,\d x
=\sum_{i,j}p_ip_j(\de_{i,j}^4-6\de_{i,j}^2+3)
	\phi(\de_{i,j})/\sigma_{i,j}^5, \cr
R_{\rm new}&=\int(f_0r'')^2\,\d x
=T_1+\cdots+T_6, \cr}$$
with these terms being defined in equation (A.1) below. 
The $R_{\rm trad}$ result is also 
proved in Marron \& Wand (1992, Theorem 4.1). 
\smallskip}}

{\csc Proof:} 
Start out noting that  
$$\int\phi_{\sigma_i}(x-\mu_i)\phi_{\sigma_j}(x-\mu_j)\,\d x
	=\phi(\de_{i,j})/\sigma_{i,j}=A_{i,j}(\mu_i,\mu_j), $$
say.
Taking derivatives with respect to $\mu_i$ and $\mu_j$ gives in 
general that 
$$\int H_r\Bigl({x-\mu_i\over \sigma_i}\Bigr)
       H_s\Bigl({x-\mu_j\over \sigma_j}\Bigr)f_i(x)f_j(x)\,\d x
	=\sigma_i^r\sigma_j^s{\dell^{r+s}\over \dell\mu_i^r\dell\mu_j^s}
	A_{i,j}=\sigma_i^r\sigma_j^sA_{i,j}^{r,s}, $$
say, $H_r$ and $H_s$ again being the Hermite polynomials. 
This leads to 
$$R_{\rm trad}(f)=\sum_{i,j}p_ip_j\int\phi_{\sigma_i}''(x-\mu_i)
	\phi_{\sigma_j}''(x-\mu_j)\,\d x
=\sum_{i,j}p_ip_j\phi^{(4)}(\de_{i,j})/\sigma_{i,j}^5, $$
proving the first and simplest assertion. 
To find $\int(f_0r'')^2\,\d x$, write (4.5) as 
$$f_0(x)r''(x)=\sum_{i=1}^kp_if_i(x)
	\{c_i+d_i(x-\mu_i)+a_i^2(x-\mu_i)^2\}, $$
where $a_i=1/\sigma_i^2-1/\sigma_0^2$,
$b_i=(\mu_i-\mu_0)/\sigma_0^2$, 
$c_i=b_i^2-a_i$, 
and $d_i=-2a_ib_i$. 
Somewhat strenuous calculations yield in the end 
the sought-for six-term expression $T_1+\cdots+T_6$ for $R_{\rm new}(f)$, 
where 
$$\eqalign{
T_1&=\sum_{i,j}p_ip_jc_ic_jA_{i,j}^{0,0}, \cr
T_2&=2\sum_{i,j}p_ip_jc_id_j\sigma_j^2A_{i,j}^{0,1}, \cr
T_3&=2\sum_{i,j}p_ip_jc_ia_j^2(\sigma_j^4A_{i,j}^{0,2}
	+\sigma_j^2A_{i,j}^{0,0}), \cr 
T_4&=\sum_{i,j}p_ip_jd_id_j\sigma_i^2\sigma_j^2A_{i,j}^{1,1}, \cr
T_5&=2\sum_{i,j}p_ip_jd_ia_j^2(\sigma_i^2\sigma_j^4A_{i,j}^{2,1}
	+\sigma_i^2\sigma_j^2A_{i,j}^{1,0}), \cr
T_6&=\sum_{i,j}p_ip_ja_i^2a_j^2\sigma_i^2\sigma_j^2
	(\sigma_i^2\sigma_j^2A_{i,j}^{2,2}+\sigma_i^2A_{i,j}^{2,0}
+\sigma_j^2A_{i,j}^{0,2}+A_{i,j}^{0,0}). \cr}\eqno({\rm A.1})$$
It is furthermore the case that 
$A_{i,j}^{r,s}=(-1)^r\phi^{(r+s)}(\delta_{i,j})/\sigma_{i,j}^{r+s+1}$. 
Hence 
$$\eqalign{
A_{i,j}^{0,0}&=\phi(\de_{i,j})/\sigma_{i,j}, \cr
A_{i,j}^{1,0}&=\de_{i,j}\phi(\de_{i,j})/\sigma_{i,j}^2=-A_{i,j}^{0,1}, \cr
A_{i,j}^{2,0}&=(\de_{i,j}^2-1)\phi(\de_{i,j})/\sigma_{i,j}^3
	=A_{i,j}^{0,2}=-A_{i,j}^{1,1}, \cr 
A_{i,j}^{2,1}&=(\de_{i,j}^3-3\de_{i,j})\phi(\de_{i,j})/\sigma_{i,j}^4
	=-A_{i,j}^{1,2}, \cr 
A_{i,j}^{2,2}&=(\de_{i,j}^4-6\de_{i,j}^2+3)
	\phi(\de_{i,j})/\sigma_{i,j}^5. \cr}$$ 
This delivers a programmable formula for $R_{\rm new}$ 
and proves the second assertion. \square 

\smallskip
In the table below we have chosen to display 
$$\rho_{\rm trad}(f)=\sigma(f)R_{\rm trad}(f)^{1/5}
\qquad {\rm and} \qquad \rho_{\rm new}(f)
	=\sigma(f)R_{\rm new}(f)^{1/5} \eqno({\rm A.2})$$
rather than $R_{\rm trad}$ and $R_{\rm new}$,
for the 15 test cases chosen in Marron \& Wand (1992). 
The $R_{\rm trad}$ values in raw form range wildly from 0.212 
to 70730, for example, and are not easily interpretable. 
The $\rho$-numbers are scale invariant 
and are directly tied to the best possible approximate mise;
the minimum amise for $\hatt f$ can be derived from (4.1) and is 
${5\over4}\sigma(f)^{-1}\{\sigma_KR(K)\}^{4/5}\rho_{\rm new}(f)/n^{4/5}$, 
with a similar expression for $\tilda f$. 

We have also included similar `difficulty measures' 
based on integrated absolute bias plus 
integrated mean absolute deviation. This is a statistically 
meaningful criterion which is also a simple upper bound on 
the expected $L_1$-distance. 
The parallel to (4.1) can be shown to be 
$$\eqalign{
{\rm (iab+imad)}(\tilda f)&
\doteq\half\sigma_K^2\int|f''|\,\d x+(2/\pi)^{1/2}R(K)^{1/2}
	(nh)^{-1/2}\int f^{1/2}\,\d x, \cr
{\rm (iab+imad)}(\hatt f)&
\doteq\half\sigma_K^2\int|f_0r''|\,\d x+(2/\pi)^{1/2}R(K)^{1/2}
	(nh)^{-1/2}\int f^{1/2}\,\d x, \cr}$$
so the values to compute and compare are primarily 
$\int|f_0r''|\,\d x$ and $\int|f''|\,\d x$. We have carried out  
numerical integrations to obtain these numbers, 
again for each of the 15 test cases. Displayed in the table are 
$$\rho^1_{\rm trad}(f)=\Bigl(\int f^{1/2}\Bigr)^{4/5}
\Bigl(\int |f''|\Bigr)^{1/5}
\quad {\rm and} \quad \rho^1_{\rm new}(f)
=\Bigl(\int f^{1/2}\Bigr)^{4/5}
	\Bigl(\int |f_0r''|\Bigr)^{1/5}. \eqno({\rm A.3})$$
This is because the minimal possible value of iab + imad for $\hatt f$
can be shown to be  
${5\over4}(2^3/\pi^2)^{1/5}\{\sigma_KR(K)\}^{2/5}\rho^1_{\rm new}(f)/n^{2/5}$,
and similarly with $\tilda f$. The quantities in 
(A.3) are scale invariant. 
\eject 

\def\qq{\qquad}
\def\ha{\hskip0.8cm}
\def\hb{\hskip1.2cm}
\def\hc{\hskip1.5cm}
\def\hd{\hskip1.2cm}

{{\medskip\narrower\noindent\sl\baselineskip11pt  
{\csc Table A.1.} 
Values of the global mise-based comparison values $\rho_{\rm trad}$ 
and $\rho_{\rm new}$, given for each of the 15 normal mixture 
test cases. Also included are the $L_1$-based global 
comparison values $\rho^1_{\rm trad}$ and $\rho^1_{\rm new}$.
The normal-start estimator (3.5) wins in approximate mise 
over the kernel method for all cases except \#12, 14, 15, where 
it loses very slightly. In terms of approximate iab plus imad 
it wins in all cases except \#3.  
\smallskip}} 

{{\smallskip\obeylines\baselineskip11pt\tt

\qq {\sl Case} \ha $\rho_{\rm trad}$ \hb $\rho_{\rm new}$ \hc $\rho^1_{\rm trad}$ \hd $\rho^1_{\rm new}$  
\smallskip
\qq ~1~~~~~~0.7330~~~~~0~~~~~~~~~~~~1.8933~~~~~0~~~~~
\qq ~2~~~~~~0.8921~~~~~0.6739~~~~~~~2.0343~~~~~1.7910
\qq ~3~~~~~~5.6070~~~~~5.5985~~~~~~~3.4988~~~~~3.5202
\qq ~4~~~~~~3.8664~~~~~3.8354~~~~~~~3.5512~~~~~3.5369
\qq ~5~~~~~~2.3201~~~~~2.2088~~~~~~~2.9388~~~~~2.9042

\qq ~6~~~~~~1.1183~~~~~1.0615~~~~~~~2.1786~~~~~2.0575
\qq ~7~~~~~~2.0215~~~~~1.9579~~~~~~~2.4701~~~~~2.4177
\qq ~8~~~~~~1.3753~~~~~1.3468~~~~~~~2.3095~~~~~2.1998
\qq ~9~~~~~~1.5600~~~~~1.5335~~~~~~~2.4608~~~~~2.3763
\qq 10~~~~~~3.5571~~~~~3.5421~~~~~~~3.8812~~~~~3.8674

\qq 11~~~~~12.4450~~~~12.4447~~~~~~~5.5611~~~~~5.5590
\qq 12~~~~~~6.4350~~~~~6.4382~~~~~~~4.0978~~~~~4.0909
\qq 13~~~~~11.1149~~~~11.1147~~~~~~~4.9481~~~~~4.9465
\qq 14~~~~~14.6610~~~~14.6615~~~~~~~4.8733~~~~~4.8703
\qq 15~~~~~~9.6259~~~~ 9.6261~~~~~~~4.3863~~~~~4.3821 
\smallskip}}



\bigskip
{\bf Appendix II: Exact mise comparisons.} 
The task considered in the following is that of 
computing the exact ${\rm mise}(h)$ for the (3.5) estimator, 
for normal mixtures. The point is to facilitate comparison 
with the kernel method, for which exact mise-calculations are known
(and much easier). 
 

Suppose again that $f(x)=\sum_{i=1}^kp_i\phi_{\sigma_i}(x-\mu_i)$
is a normal mixture. Start out with 
$${\rm ise}(h)=\int(\hatt f-f)^2\,\d x
	=A_h-2B_h+R(f), \eqno({\rm A.4})$$ 
where  
$$R(f)=\int f^2\,\d x=\sum_{i,j}p_ip_j\phi\Bigl({\mu_j-\mu_i
	\over \sigma_{i,j}}\Bigr){1\over \sigma_{i,j}}, \eqno({\rm A.5})$$
again using $\sigma_{i,j}=(\sigma_i^2+\sigma_j^2)^{1/2}$.
To give useful expressions for $A_h$ and $B_h$ we note the
technical fact that  
$$\int\prod_{j=1}^m\phi_{\sigma_j}(x-\mu_j)\,\d x
=\sqrt{2\pi}\tilda\sigma\Bigl[\prod_{j=1}^m\phi_{\sigma_j}(\mu_j-a)\Bigr]
\exp\Bigl[\half\tilda\sigma^2
\Bigl\{\sum_{j=1}^m(\mu_j-a)/\sigma_j^2\Bigr\}^2\Bigr], \eqno({\rm A.6})$$
where $1/\tilda\sigma^2=\sum_{j=1}^m1/\sigma_j^2$. 
The value of $a$ is arbitrary and can be chosen for the occasion. 
Proving (A.6) is not very difficult and we omit the details. 
For the first term this identity gives 
$$\eqalign{A_h
&={1\over n^2}\sum_{i,j\le n}\int
{\phi_h(x-x_i)\phi_h(x-x_j)\phi_{\hatt\sigma}(x-\hatt\mu)^2
\over \phi_{\hatt\sigma}(x_i-\hatt\mu)\phi_{\hatt\sigma}(x_j-\hatt\mu)} 
	\,\d x \cr
&={1\over n^2}\sum_{i,j\le n}
{1\over 2\rootpi}{\tilda\sigma\over \hatt\sigma^2}
{\phi_h(x_i-\hatt\mu)\over \phi_{\hatt\sigma}(x_i-\hatt\mu)}
{\phi_h(x_j-\hatt\mu)\over \phi_{\hatt\sigma}(x_j-\hatt\mu)}	
	\exp\Bigl\{\half\tilda\sigma^2
\Bigl({x_i-\hatt\mu+x_j-\hatt\mu\over h^2}\Bigr)^2\Bigr\}, \cr}$$
where 
$$\tilda\sigma^2=\Bigl({2\over \hatt\sigma^2}+{2\over h^2}\Bigr)^{-1}
	=\half{\hatt\sigma^2\over \hatt\sigma^2+h^2}h^2. $$
And for the second term, 
$$\eqalign{B_h
&=\sum_{j=1}^kp_j{1\over n}\sumin\int\phi_h(x-x_i)
{\phi_{\hatt\sigma}(x-\hatt\mu)\over \phi_{\hatt\sigma}(x_i-\hatt\mu)}
	\phi_{\sigma_j}(x-\mu_j)\,\d x \cr
&={1\over \hatt\sigma}
\sum_{j=1}^k p_j\tilda\sigma_j\phi_{\sigma_j}(\mu_j-\hatt\mu)
\Bigl[{1\over n}\sumin{\phi_h(x_i-\hatt\mu)
	\over \phi_{\hatt\sigma}(x_i-\hatt\mu)}
\exp\Bigl\{\half\bar\sigma_j^2\Bigl({x_i-\hatt\mu\over h^2}
+{\mu_j-\hatt\mu\over \sigma_j^2}\Bigr)^2\Bigr\}\Bigr], \cr}$$
where this time 
$$\tilda\sigma_j^2=\Bigl({1\over h^2}+{1\over \hatt\sigma^2}
	+{1\over \sigma_j^2}\Bigr)^{-1}
={\hatt\sigma^2\sigma_j^2\over \hatt\sigma^2\sigma_j^2
	+h^2(\hatt\sigma^2+\sigma_j^2)}h^2. $$

Finding further exact expressions for the mise involves 
finding the exact means of $A_h$ and $B_h$.
This would depend on the parameter estimation method used,
and in any case seems forbiddingly difficult.  
The formulae for $A_h$ and $B_h$ can however be used to
compute their mean values, and hence the mise, via stochastic simulation,
for each given mixture and each given sample size. 
At this stage we are content to find the exact mise 
for the estimator that employs true parameter values 
$\mu_0$ and $\sigma_0$ for mean and standard deviation. 
This allows a `best case versus best case' comparison 
with the kernel method to be made, and the extra 
variability caused by using parameter estimates for 
$\mu$ and $\sigma$ is in any case of second order importance.  

{{\smallskip\sl
{\csc Proposition A.2.} 
Consider the normal start times correction estimator
with the normal kernel,  
$$\hatt f(x)={1\over n}\sumin \phi_h(X_i-x)
{\phi_{\sigma_0}(x-\mu_0)\over \phi_{\sigma_0}(X_i-\mu_0)}
={1\over n}\sumin \phi_h(X_i-x)
{\exp\{-\half(x-\mu_0)^2/\sigma_0^2\}
\over \exp\{-\half(X_i-\mu_0)^2/\sigma_0^2\}}. $$
Its exact mean integrated squared error,
in the case when $f(x)=\sum_{i=1}^kp_i\phi_{\sigma_i}(x-\mu_i)$
is a normal mixture, can be expressed as 
$${\rm mise}(h)=(1-n^{-1})\,\E A_{1,h}+n^{-1}\,\E A_{2,h}
	-2\,\E B_h+R(f), \eqno({\rm A.7})$$
where $R(f)$ is given in (A.5) and where formulae for 
the other three terms appear in equations (A.8--10) below.
\smallskip}}

{\csc Proof:} 
We start out finding an exact expression for the expected value:
$$\eqalign{\E\hatt f(x)
&=\int\phi_h(y-x){\phi_{\sigma_0}(x-\mu_0)\over \phi_{\sigma_0}(y-\mu_0)}
	f(y)\,\d y \cr
&=\phi_{\sigma_0}(x-\mu_0)\sum_{i=1}^k p_i
\int\phi(z){\phi_{\sigma_i}(x-\mu_i+hz)
	\over \phi_{\sigma_0}(x-\mu_0+hz)}\,\d z, \cr}$$
using the $z=(y-x)/h$ substitution. Expanding the exponent 
and collecting $z^2$ terms, and using 
$$b_i=\Bigl(1+{h^2\over \sigma_i^2}-{h^2\over \sigma_0^2}\Bigr)^{1/2}, $$
one finds 
$$\E\hatt f(x)={1\over \sqrt{2\pi}}\sum_{i=1}^kp_i{1\over \sigma_ib_i}
\exp\Bigl\{-\half{(x-\mu_i)^2\over \sigma_i^2}
	+\half\Bigl({x-\mu_i\over \sigma_i^2}
	-{x-\mu_0\over \sigma_0^2}\Bigr)^2{h^2\over b_i^2}\Bigr\}. $$
Indeed this is $f(x)+O(h^2)$. 
Next consider $A_h$ of (A.4) and its mean value. Splitting $A_h$ into 
non-diagonal and diagonal terms leads to 
$\E A_h=(1-n^{-1})\,\E A_{1,h}+n^{-1}\,\E A_{2,h}$,
leaving us the task of calculating $\E A_{1,h}$ and $\E A_{2,h}$
by integration. First, 
$$\eqalign{\E A_{1,h}
&=\E\int\Bigl\{
{\phi_h(x-X_1)\phi_{\sigma_0}(x-\mu_0)
	\over \phi_{\sigma_0}(X_1-\mu_0)}
{\phi_h(x-X_2)\phi_{\sigma_0}(x-\mu_0)
	\over \phi_{\sigma_0}(X_2-\mu_0)}\Bigr\}\,\d x \cr
&=\int\{\E\hatt f(x)\}^2\,\d x \cr
&={1\over 2\pi}\sum_{i,j}p_ip_j{1\over \sigma_i\sigma_jb_ib_j}
\int\exp\Bigl\{-\half{(x-\mu_i)^2\over \sigma_i^2}
-\half{(x-\mu_j)^2\over \sigma_j^2} \cr 
&\qquad\qquad 
+\half\Bigl({x-\mu_i\over \sigma_i^2}
	-{x-\mu_0\over \sigma_0^2}\Bigr)^2{h^2\over b_i^2}
+\half\Bigl({x-\mu_j\over \sigma_j^2}
	-{x-\mu_0\over \sigma_0^2}\Bigr)^2{h^2\over b_j^2}\Bigr\}. \cr}$$
Collecting $x^2$ terms and transforming to the standard normal,
employing 
$$\eqalign{
c_{i,j}&=\Bigl\{{1\over \sigma_i^2}+{1\over \sigma_j^2}
-\Bigl({1\over \sigma_i^2}-{1\over \sigma_0^2}\Bigr)^2
	{h^2\over b_i^2}
-\Bigl({1\over \sigma_j^2}-{1\over \sigma_0^2}\Bigr)^2
	{h^2\over b_j^2}\Bigr\}^{1/2}, \cr
d_{i,j}&={\mu_i\over \sigma_i^2}+{\mu_j\over \sigma_j^2}
-\Bigl({1\over \sigma_i^2}-{1\over \sigma_0^2}\Bigr)
 \Bigl({\mu_i\over \sigma_i^2}-{\mu_0\over \sigma_0^2}\Bigr){h^2\over b_i^2}
-\Bigl({1\over \sigma_j^2}-{1\over \sigma_0^2}\Bigr)
 \Bigl({\mu_j\over \sigma_j^2}-{\mu_0\over \sigma_0^2}\Bigr){h^2\over b_j^2},
	\cr}$$
the result is 
$$\eqalign{\E A_{1,h}
&=\sqrt{2\pi}\sum_{i,j}{p_ip_j\over b_ib_j}
\phi_{\sigma_i}(\mu_i)\phi_{\sigma_j}(\mu_j)
{1\over c_{i,j}}\exp\Bigl\{\half{d_{i,j}^2\over c_{i,j}^2} \cr 
&\qquad\qquad\qquad\qquad 
+\half\Bigl({\mu_i\over \sigma_i^2}-{\mu_0\over \sigma_0^2}\Bigr)^2
	{h^2\over b_i^2}	
+\half\Bigl({\mu_j\over \sigma_j^2}-{\mu_0\over \sigma_0^2}\Bigr)^2
	{h^2\over b_j^2}\Bigr\}. \cr} \eqno({\rm A.8})$$	
Similar and somewhat arduous calculations yield the mean of $A_{2,h}$.
The starting point is 
$$\eqalign{\E A_{2,h}
&=\E\int\Bigl\{\phi_{\sigma_0}(x-\mu_0)
	{\phi_h(X_i-x)\over \phi_{\sigma_0}(X_i-\mu_0)}\Bigr\}^2\,\d x \cr
&=\int\phi_{\sigma_0}(x-\mu_0)^2\Bigl\{\sum_{i=1}^kp_i
\int{\phi_h(y-x)^2\over \phi_{\sigma_0}(y-\mu_0)^2}
	\phi_{\sigma_i}(y-\mu_i)\,\d y\Bigr\}\,\d x \cr
&=h^{-1}\sum_{i=1}^kp_i
\int\phi_{\sigma_0}(x-\mu_0)^2\Bigl\{\int\phi(z)^2
{\phi_{\sigma_i}(x-\mu_i+hz)\over \phi_{\sigma_0}(x-\mu_0+hz)^2}\,\d z
	\Bigr\}\,\d x. \cr}$$
Again $z^2$ terms have to be collected for the inner integral and 
then $x^2$ terms to do the rest. We need to introduce 
$$\eqalign{
e_i&=\Bigl(2+{h^2\over \sigma_i^2}-2{h^2\over \sigma_0^2}\Bigr)^{1/2}, \cr
f_i&=\Bigl\{{1\over \sigma_i^2}
-\Bigl({1\over \sigma_i^2}-{2\over \sigma_0^2}\Bigr)^2{h^2\over e_i^2}
	\Bigr\}^{1/2}, \cr
g_i&={\mu_i\over \sigma_i^2}
-\Bigl({1\over \sigma_i^2}-{2\over \sigma_0^2}\Bigr)
\Bigl({\mu_i\over \sigma_0^2}-2{\mu_0\over \sigma_0^2}\Bigr)
	{h^2\over e_i^2}. \cr}$$
The answer is 
$$\E A_{2,h}={h^{-1}\over \sqrt{2\pi}}
\sum_{i=1}^k{p_i\over \sigma_ie_if_i}
\exp\Bigl\{\half{g_i^2\over f_i^2}-\half{\mu_i^2\over \sigma_i^2}
+\half\Bigl({\mu_i\over \sigma_i^2}-2{\mu_0\over \sigma_0^2}\Bigr)^2
	{h^2\over e_i^2}\Bigr\}. \eqno({\rm A.9})$$
This is close to $h^{-1}(2\sqrt{\pi})^{-1}$ when $h$ is small. 

It remains only to find the mean of $B_h=\int f\hatt f\,\d x$.
By our earlier result about the exact mean of $\hatt f$ this is equal to
$$\eqalign{\E B_h
&=\int f(x)\,\E\hatt f(x)\,\d x \cr
&={1\over \sqrt{2\pi}}\sum_{i,j}p_ip_j{1\over \sigma_ib_i}
\int\exp\Bigl\{\half\Bigl({x-\mu_i\over \sigma_i^2}
-{x-\mu_0\over \sigma_0^2}\Bigr)^2{h^2\over b_i^2} \cr
&\qquad\qquad\qquad\qquad 
-\half{(x-\mu_i)^2\over \sigma_i^2}\Bigr\}
\phi_{\sigma_j}(x-\mu_j)\,\d x. \cr}$$
This time we need 
$$\eqalign{
k_{i,j}&=\Bigl\{{1\over \sigma_i^2}+{1\over \sigma_j^2}
-\Bigl({1\over \sigma_i^2}-{1\over \sigma_0^2}\Bigr)^2{h^2\over b_i^2}
	\Bigr\}^{1/2}, \cr
l_{i,j}&={\mu_i\over \sigma_i^2}+{\mu_j\over \sigma_j^2}
-\Bigl({1\over \sigma_i^2}-{1\over \sigma_0^2}\Bigr)
 \Bigl({\mu_i\over \sigma_i^2}
	-{\mu_0\over \sigma_0^2}\Bigr){h^2\over b_i^2}, \cr}$$
and the result is 
$$\E B_h=\sqrt{2\pi}\sum_{i,j}
p_ip_j\phi_{\sigma_i}(\mu_i)\phi_{\sigma_j}(\mu_j)
{1\over b_i}{1\over k_{i,j}}
\exp\Bigl\{\half\Bigl({\mu_i\over \sigma_i^2}-{\mu_0\over \sigma_0^2}\Bigr)^2
{h^2\over b_i^2}+\half{l_{i,j}^2\over k_{i,j}^2}\Bigr\}. \eqno({\rm A.10})$$
This ends our proof. \square 

\smallskip
Consider the limiting case where $\sigma_0\arr\infty$. 
Then our estimator is nothing but the usual kernel estimator. 
Somewhat strenuous algebraic calculations yield 
$$\eqalign{
\E A_{1,h}&=\sum_{i,j}p_ip_j
	\phi_{(\sigma_i^2+\sigma_j^2+2h^2)^{1/2}}(\mu_j-\mu_i), \cr
\E A_{2,h}&=(2\sqrt{\pi}h)^{-1}, \cr
\E B_h&=\sum_{i,j}p_ip_j
	\phi_{(\sigma_i^2+\sigma_j^2+h^2)^{1/2}}(\mu_j-\mu_i), \cr}$$
which with (A.7) and (A.5) again quite satisfactorily give
the (5.4) formula for the exact ${\rm mise}(h)$ of the kernel estimator. 

\smallskip
We used these results to go through the 15 test densities 
of Marron \& Wand (1992), with the natural aim of comparing 
the minimum possible mise for the kernel method with
the minimum possible mise for the new method (3.5). 
These minima, respectively ${\rm mise}_{\rm trad}^*$ and ${\rm mise}^*$,
were found, along with the minimisers $h^*_{\rm trad}$ and $h^*$,   
for sample sizes $n=25,50,100,250,1000$. 
See the discussion of Section 5B. 

{{\medskip\narrower\noindent\sl\baselineskip11pt  
{\csc Table A.2.} 
Values are given of the mise-minimising
smoothing parameters $h^*$ and $h^*_{\rm trad}$
for the (3.5) estimator and the kernel estimator, 
along with the minimum mise values 
${\rm mise}^*$ and ${\rm mise}^*_{\rm trad}$. 
This is done for each of the 15 test densities of Marron \& Wand,
for sample sizes 25, 50, 100, 200, 1000. 
Also included in each case is the ratio 
${\rm mise}^*/{\rm mise}_{\rm trad}^*$.
\smallskip}} 

\def\qq{\qquad}
\def\ha{\hskip0.8cm}
\def\hb{\hskip1.0cm}
\def\hc{\hskip0.8cm}
\def\hd{\hskip0.7cm}
\def\he{\hskip0.4cm}
\def\hf{\hskip0.20cm}
{{\smallskip\obeylines\tt\baselineskip11pt 
    \ha $n$ \hb $h^*$ \hc ${\rm mise}^*$ \hd $h^*_{\rm trad}$
    \he ${\rm mise}^*_{\rm trad}$ \hf {\rm mise-ratio}

\smallskip
\qq {\rm Case \#1, Gau\ss ian:}       
~~~ 25~~~0.7071~~~0.0113~~~0.6094~~~0.0137~~~0.8217
~~~ 50~~~0.7071~~~0.0056~~~0.5199~~~0.0087~~~0.6492
~~~100~~~0.7071~~~0.0028~~~0.4455~~~0.0054~~~0.5215
~~~200~~~0.7071~~~0.0014~~~0.3830~~~0.0033~~~0.4245
~~1000~~~0.7071~~~0.0003~~~0.2723~~~0.0010~~~0.2740

\smallskip
\qq {\rm Case \#2, skewed unimodal:} 
~~~ 25~~~0.3928~~~0.0228~~~0.4251~~~0.0211~~~1.0772
~~~ 50~~~0.3787~~~0.0123~~~0.3591~~~0.0134~~~0.9173
~~~100~~~0.3544~~~0.0068~~~0.3054~~~0.0083~~~0.8250
~~~200~~~0.3209~~~0.0040~~~0.2611~~~0.0051~~~0.7767
~~1000~~~0.2381~~~0.0012~~~0.1841~~~0.0016~~~0.7396

\smallskip
\qq {\rm Case \#3, strongly skewed:} 
~~~ 25~~~0.0728~~~0.1456~~~0.1481~~~0.1032~~~1.4107
~~~ 50~~~0.0720~~~0.0786~~~0.1082~~~0.0682~~~1.1523
~~~100~~~0.0720~~~0.0444~~~0.0827~~~0.0435~~~1.0208
~~~200~~~0.0655~~~0.0270~~~0.0654~~~0.0270~~~0.9996
~~1000~~~0.0415~~~0.0084~~~0.0414~~~0.0084~~~0.9989

\smallskip
\qq {\rm Case \#4, kurtotic unimodal:} 
~~~ 25~~~0.1252~~~0.1098~~~0.1241~~~0.1101~~~0.9972
~~~ 50~~~0.0976~~~0.0688~~~0.0967~~~0.0691~~~0.9949
~~~100~~~0.0791~~~0.0421~~~0.0784~~~0.0424~~~0.9937
~~~200~~~0.0656~~~0.0253~~~0.0650~~~0.0255~~~0.9930
~~1000~~~0.0445~~~0.0075~~~0.0441~~~0.0076~~~0.9922

\smallskip
\qq {\rm Case \#5, outlier:} 
~~~ 25~~~0.0634~~~0.1433~~~0.0646~~~0.1424~~~1.0062
~~~ 50~~~0.0562~~~0.0862~~~0.0548~~~0.0890~~~0.9690
~~~100~~~0.0487~~~0.0523~~~0.0468~~~0.0548~~~0.9549
~~~200~~~0.0420~~~0.0317~~~0.0402~~~0.0334~~~0.9492
~~1000~~~0.0299~~~0.0096~~~0.0285~~~0.0102~~~0.9462 

\smallskip
\qq {\rm Case \#6, bimodal:} 
~~~ 25~~~0.5568~~~0.0197~~~0.6028~~~0.0182~~~1.0792
~~~ 50~~~0.4559~~~0.0123~~~0.4721~~~0.0119~~~1.0342
~~~100~~~0.3823~~~0.0075~~~0.3854~~~0.0075~~~1.0067
~~~200~~~0.3247~~~0.0045~~~0.3217~~~0.0046~~~0.9888
~~1000~~~0.2278~~~0.0013~~~0.2208~~~0.0014~~~0.9663

\smallskip
\qq {\rm Case \#7, separated bimodal:} 
~~~ 25~~~0.3701~~~0.0303~~~0.3661~~~0.0306~~~0.9881
~~~ 50~~~0.3136~~~0.0183~~~0.3082~~~0.0187~~~0.9813
~~~100~~~0.2674~~~0.0110~~~0.2616~~~0.0112~~~0.9768
~~~200~~~0.2291~~~0.0065~~~0.2235~~~0.0067~~~0.9738
~~1000~~~0.1620~~~0.0019~~~0.1575~~~0.0020~~~0.9700

\smallskip
\qq {\rm Case \#8, skewed bimodal:} 
~~~ 25~~~0.5136~~~0.0243~~~0.5549~~~0.0222~~~1.0953
~~~ 50~~~0.3903~~~0.0158~~~0.4085~~~0.0151~~~1.0507
~~~100~~~0.3112~~~0.0100~~~0.3179~~~0.0097~~~1.0251
~~~200~~~0.2554~~~0.0061~~~0.2572~~~0.0061~~~1.0099
~~1000~~~0.1712~~~0.0019~~~0.1697~~~0.0019~~~0.9924

\smallskip
\qq {\rm Case \#9, trimodal:} 
~~~ 25~~~0.5373~~~0.0224~~~0.5889~~~0.0206~~~1.0840
~~~ 50~~~0.4331~~~0.0144~~~0.4551~~~0.0138~~~1.0435
~~~100~~~0.3509~~~0.0091~~~0.3588~~~0.0089~~~1.0193
~~~200~~~0.2858~~~0.0057~~~0.2874~~~0.0056~~~1.0052
~~1000~~~0.1848~~~0.0018~~~0.1829~~~0.0018~~~0.9910

\smallskip
\qq {\rm Case \#10, claw:} 
~~~ 25~~~0.4930~~~0.0659~~~0.5101~~~0.0636~~~1.0372
~~~ 50~~~0.4267~~~0.0578~~~0.4034~~~0.0570~~~1.0145
~~~100~~~0.0955~~~0.0371~~~0.0959~~~0.0370~~~1.0033
~~~200~~~0.0774~~~0.0224~~~0.0775~~~0.0224~~~1.0007
~~1000~~~0.0517~~~0.0067~~~0.0516~~~0.0067~~~0.9979

\smallskip
\qq {\rm Case \#11, double claw:} 
~~~ 25~~~0.5556~~~0.0212~~~0.6018~~~0.0197~~~1.0748
~~~ 50~~~0.4550~~~0.0138~~~0.4717~~~0.0134~~~1.0318
~~~100~~~0.3817~~~0.0090~~~0.3851~~~0.0089~~~1.0073
~~~200~~~0.3242~~~0.0060~~~0.3215~~~0.0061~~~0.9925
~~1000~~~0.2248~~~0.0028~~~0.2176~~~0.0029~~~0.9854

\smallskip
\qq {\rm Case \#12, asymmetric claw:} 
~~~ 25~~~0.7289~~~0.0363~~~0.6657~~~0.0359~~~1.0121
~~~ 50~~~0.6044~~~0.0312~~~0.5231~~~0.0309~~~1.0079
~~~100~~~0.1989~~~0.0232~~~0.2016~~~0.0229~~~1.0115
~~~200~~~0.1428~~~0.0161~~~0.1436~~~0.0160~~~1.0073
~~1000~~~0.0675~~~0.0064~~~0.0678~~~0.0064~~~1.0043

\smallskip
\qq {\rm Case \#13, asymmetric double claw:} 
~~~ 25~~~0.5254~~~0.0254~~~0.5620~~~0.0241~~~1.0532
~~~ 50~~~0.4315~~~0.0174~~~0.4428~~~0.0171~~~1.0188
~~~100~~~0.3608~~~0.0123~~~0.3612~~~0.0123~~~1.0008
~~~200~~~0.3021~~~0.0091~~~0.2971~~~0.0091~~~0.9937
~~1000~~~0.1030~~~0.0045~~~0.1030~~~0.0045~~~1.0010

\smallskip
\qq {\rm Case \#14, smooth comb:} 
~~~ 25~~~0.2866~~~0.0678~~~0.2858~~~0.0675~~~1.0037
~~~ 50~~~0.2035~~~0.0488~~~0.2031~~~0.0487~~~1.0026
~~~100~~~0.1434~~~0.0348~~~0.1434~~~0.0347~~~1.0021
~~~200~~~0.1015~~~0.0245~~~0.1016~~~0.0244~~~1.0018
~~1000~~~0.0439~~~0.0101~~~0.0439~~~0.0101~~~1.0007

\smallskip
\qq {\rm Case \#15, discrete comb:} 
~~~ 25~~~0.2459~~~0.0704~~~0.2469~~~0.0702~~~1.0033
~~~ 50~~~0.2016~~~0.0493~~~0.2014~~~0.0493~~~1.0007
~~~100~~~0.1638~~~0.0362~~~0.1630~~~0.0362~~~0.9998
~~~200~~~0.0815~~~0.0266~~~0.0816~~~0.0266~~~1.0016
~~1000~~~0.0422~~~0.0087~~~0.0423~~~0.0087~~~1.0006

\smallskip}}

\bigskip
\baselineskip11pt
{\bf Acknowledgements.} 
We are grateful for useful and encouraging comments from M.C.~Jones
and for discussions with Grete Fenstad. 
A part of this work was carried out while one of was
visiting Oxford University 
with a grant from the Royal Norwegian Research Council.   

\bigskip
 
\parindent0pt
\parskip3pt
  
\centerline{\bf References}

\medskip



\ref{%
Buckland, S.T. (1992).
Maximum likelihood fitting of Hermite and simple polynomial densities.
{\sl Applied Statistics} {\bf 41}, 241--266.}

\ref{%
Friedman, J.H., Stuetzle, W., and Schroeder, A. (1984).
Projection pursuit density estimation.
{\sl Journal of the American Statistical Association} {\bf 79}, 599--608.}


\ref{%
Hall, P.G., Sheather, S.J., Jones., M.C., and Marron, S.J. (1991).
On optimal data-based bandwidth selection in kernel density estimation.
{\sl Biometrika} {\bf 78}, 263--269.}

\ref{%
Hjort, N.L. (1986). 
{\sl Statistical Symbol Recognition.}
Research monograph, Norwegian Computing Centre, Oslo.}


\ref{%
Hjort, N.L. (1993).
Dynamic likelihood hazard rate estimation.
{\sl Biometrika}, to appear.}


\ref{%
Hjort, N.L. (1994).
Bayesian approaches to semiparametric density estimation.
Invited paper, in progress, 
to be published in the proceedings of the 
Fifth Valencia International Meeting on Bayesian Statistics.}

\ref{%
Hjort, N.L.~and Jones, M.C. (1993).
Locally parametric nonparametric density estimation.
Statistical Research Report, Department of Mathematics,
University of Oslo. Submitted for publication.}

\ref{%
Hjort, N.L.~and Jones, M.C. (1994).
Better rules of thumb for choice of smoothing parameter in 
density estimation. 
In progress.}

\ref{%
Hjort, N.L.~and Fenstad, G.U. (1994).
Hermite versus Kernel. 
In progress.}


\ref{%
Huber, P.J. (1981).
{\sl Robust Statistics.}
Wiley, New York.}

\ref{%
Jones, M.C. (1993).
Kernel density estimation when the bandwidth is large.
{\sl Australian Journal of Statistics}, to appear.}

\ref{%
Jones, M.C., Linton, O.~and Nielsen, J.P. (1993).
A simple and effective bias reduction method for density and regression
estimation. Manuscript.}

\ref{%
Jones, M.C., Marron, J.S.~and Sheather, S.J. (1993).
Progress in data-based bandwidth selection for kernel density estimation. 
Working Paper 92--014, Australian Graduate School of Management,
University of New South Wales.}

\ref{%
Marron, J.S.~and Wand, M.P. (1992).
Exact mean integrated squared error.
{\sl Annals of Statistics} {\bf 20}, 712--736.}

\ref{%
Olkin, I.~and Spiegelman, C.H. (1987).
A semiparametric approach to density estimation.
{\sl Journal of the American Statistical Association} {\bf 82}, 858--865.}


\ref{%
Schuster, E. and Yakowitz, S. (1985).
Parametric/nonparametric mixture density estimation with application to
flood-frequency analysis.
{\sl Water Resources Bulletin} {\bf 21}, 797--804.}

\ref{%
Scott, D.W. (1992).
{\sl Multivariate Density Estimation:
Theory, Practice, and Visualization.}
Wiley, New York.}

\ref{%
Scott, D.W.~and Terrell, G.R. (1987).
Biased and unbiased cross-validation in density estimation.
{\sl Journal of the American Statistical Association} {\bf 82}, 1131--1146.}

\ref{%
Shao, J. (1991). 
Second-order differentiability and jackknife.
{\sl Statistica Sinica} {\bf 1}, 185--202.}

\ref{%
Sheather, S.J.~and Jones, M.C. (1991).
A reliable data-based bandwidth selection method for kernel
density estimation.
{\sl Journal of the Royal Statistical Association}~{\bf B 53}, 683--690.}

\ref{%
Wand, M.P.~and Jones, M.C. (1994).
{\sl Kernel Smoothing.}
Chapman \& Hall, London. To exist.}  

\ref{%
Wand, M.P., Marron, J.S., and Ruppert, D. (1991).
Transformations in density estimation [with discussion contributions]. 
{\sl Journal of the American Statistical Association} {\bf 86},
343--361.}


\bye

~ 

\nopagenumbers

\vskip21.5truecm 

{{\medskip\narrower\noindent\baselineskip11pt\sl
{\csc Figure.}
\fermat{This figure belongs to the discussion of Section 5A}The 
15 test densities (left hand side) presented together with the bias factor 
functions $f''$ (solid line, for the kernel method) and $f_0r''$ 
(dotted line, for the new method). 
\smallskip}}
\eject 

\bye